\documentclass[aps,prd,twocolumn,superscriptaddress,showpacs]{revtex4}

\usepackage{epsfig}
\usepackage{lscape}
\usepackage{graphicx}
\usepackage{dcolumn}
\usepackage{bm}
\usepackage{setspace}

\begin{document}
\renewcommand{\thetable}{\Roman{table}}

\title{Mass and lifetime measurements of bottom and charm baryons 
in $p\bar p$ collisions at $\sqrt{s}~=$~1.96~TeV}
\affiliation{Institute of Physics, Academia Sinica, Taipei, Taiwan 11529, Republic of China}
\affiliation{Argonne National Laboratory, Argonne, Illinois 60439, USA}
\affiliation{University of Athens, 157 71 Athens, Greece}
\affiliation{Institut de Fisica d'Altes Energies, ICREA, Universitat Autonoma de Barcelona, E-08193, Bellaterra (Barcelona), Spain}
\affiliation{Baylor University, Waco, Texas 76798, USA}
\affiliation{Istituto Nazionale di Fisica Nucleare Bologna, \ensuremath{^{ii}}University of Bologna, I-40127 Bologna, Italy}
\affiliation{University of California, Davis, Davis, California 95616, USA}
\affiliation{University of California, Los Angeles, Los Angeles, California 90024, USA}
\affiliation{Instituto de Fisica de Cantabria, CSIC-University of Cantabria, 39005 Santander, Spain}
\affiliation{Carnegie Mellon University, Pittsburgh, Pennsylvania 15213, USA}
\affiliation{Enrico Fermi Institute, University of Chicago, Chicago, Illinois 60637, USA}
\affiliation{Comenius University, 842 48 Bratislava, Slovakia; Institute of Experimental Physics, 040 01 Kosice, Slovakia}
\affiliation{Joint Institute for Nuclear Research, RU-141980 Dubna, Russia}
\affiliation{Duke University, Durham, North Carolina 27708, USA}
\affiliation{Fermi National Accelerator Laboratory, Batavia, Illinois 60510, USA}
\affiliation{University of Florida, Gainesville, Florida 32611, USA}
\affiliation{Laboratori Nazionali di Frascati, Istituto Nazionale di Fisica Nucleare, I-00044 Frascati, Italy}
\affiliation{University of Geneva, CH-1211 Geneva 4, Switzerland}
\affiliation{Glasgow University, Glasgow G12 8QQ, United Kingdom}
\affiliation{Harvard University, Cambridge, Massachusetts 02138, USA}
\affiliation{Division of High Energy Physics, Department of Physics, University of Helsinki, FIN-00014, Helsinki, Finland; Helsinki Institute of Physics, FIN-00014, Helsinki, Finland}
\affiliation{University of Illinois, Urbana, Illinois 61801, USA}
\affiliation{The Johns Hopkins University, Baltimore, Maryland 21218, USA}
\affiliation{Institut f\"{u}r Experimentelle Kernphysik, Karlsruhe Institute of Technology, D-76131 Karlsruhe, Germany}
\affiliation{Center for High Energy Physics: Kyungpook National University, Daegu 702-701, Korea; Seoul National University, Seoul 151-742, Korea; Sungkyunkwan University, Suwon 440-746, Korea; Korea Institute of Science and Technology Information, Daejeon 305-806, Korea; Chonnam National University, Gwangju 500-757, Korea; Chonbuk National University, Jeonju 561-756, Korea; Ewha Womans University, Seoul, 120-750, Korea}
\affiliation{Ernest Orlando Lawrence Berkeley National Laboratory, Berkeley, California 94720, USA}
\affiliation{University of Liverpool, Liverpool L69 7ZE, United Kingdom}
\affiliation{University College London, London WC1E 6BT, United Kingdom}
\affiliation{Centro de Investigaciones Energeticas Medioambientales y Tecnologicas, E-28040 Madrid, Spain}
\affiliation{Massachusetts Institute of Technology, Cambridge, Massachusetts 02139, USA}
\affiliation{University of Michigan, Ann Arbor, Michigan 48109, USA}
\affiliation{Michigan State University, East Lansing, Michigan 48824, USA}
\affiliation{Institution for Theoretical and Experimental Physics, ITEP, Moscow 117259, Russia}
\affiliation{University of New Mexico, Albuquerque, New Mexico 87131, USA}
\affiliation{The Ohio State University, Columbus, Ohio 43210, USA}
\affiliation{Okayama University, Okayama 700-8530, Japan}
\affiliation{Osaka City University, Osaka 558-8585, Japan}
\affiliation{University of Oxford, Oxford OX1 3RH, United Kingdom}
\affiliation{Istituto Nazionale di Fisica Nucleare, Sezione di Padova, \ensuremath{^{jj}}University of Padova, I-35131 Padova, Italy}
\affiliation{University of Pennsylvania, Philadelphia, Pennsylvania 19104, USA}
\affiliation{Istituto Nazionale di Fisica Nucleare Pisa, \ensuremath{^{kk}}University of Pisa, \ensuremath{^{ll}}University of Siena, \ensuremath{^{mm}}Scuola Normale Superiore, I-56127 Pisa, Italy, \ensuremath{^{nn}}INFN Pavia, I-27100 Pavia, Italy, \ensuremath{^{oo}}University of Pavia, I-27100 Pavia, Italy}
\affiliation{University of Pittsburgh, Pittsburgh, Pennsylvania 15260, USA}
\affiliation{Purdue University, West Lafayette, Indiana 47907, USA}
\affiliation{University of Rochester, Rochester, New York 14627, USA}
\affiliation{The Rockefeller University, New York, New York 10065, USA}
\affiliation{Istituto Nazionale di Fisica Nucleare, Sezione di Roma 1, \ensuremath{^{pp}}Sapienza Universit\`{a} di Roma, I-00185 Roma, Italy}
\affiliation{Mitchell Institute for Fundamental Physics and Astronomy, Texas A\&M University, College Station, Texas 77843, USA}
\affiliation{Istituto Nazionale di Fisica Nucleare Trieste, \ensuremath{^{qq}}Gruppo Collegato di Udine, \ensuremath{^{rr}}University of Udine, I-33100 Udine, Italy, \ensuremath{^{ss}}University of Trieste, I-34127 Trieste, Italy}
\affiliation{University of Tsukuba, Tsukuba, Ibaraki 305, Japan}
\affiliation{Tufts University, Medford, Massachusetts 02155, USA}
\affiliation{University of Virginia, Charlottesville, Virginia 22906, USA}
\affiliation{Waseda University, Tokyo 169, Japan}
\affiliation{Wayne State University, Detroit, Michigan 48201, USA}
\affiliation{University of Wisconsin, Madison, Wisconsin 53706, USA}
\affiliation{Yale University, New Haven, Connecticut 06520, USA}

\author{T.~Aaltonen}
\affiliation{Division of High Energy Physics, Department of Physics, University of Helsinki, FIN-00014, Helsinki, Finland; Helsinki Institute of Physics, FIN-00014, Helsinki, Finland}
\author{S.~Amerio\ensuremath{^{jj}}}
\affiliation{Istituto Nazionale di Fisica Nucleare, Sezione di Padova, \ensuremath{^{jj}}University of Padova, I-35131 Padova, Italy}
\author{D.~Amidei}
\affiliation{University of Michigan, Ann Arbor, Michigan 48109, USA}
\author{A.~Anastassov\ensuremath{^{v}}}
\affiliation{Fermi National Accelerator Laboratory, Batavia, Illinois 60510, USA}
\author{A.~Annovi}
\affiliation{Laboratori Nazionali di Frascati, Istituto Nazionale di Fisica Nucleare, I-00044 Frascati, Italy}
\author{J.~Antos}
\affiliation{Comenius University, 842 48 Bratislava, Slovakia; Institute of Experimental Physics, 040 01 Kosice, Slovakia}
\author{G.~Apollinari}
\affiliation{Fermi National Accelerator Laboratory, Batavia, Illinois 60510, USA}
\author{J.A.~Appel}
\affiliation{Fermi National Accelerator Laboratory, Batavia, Illinois 60510, USA}
\author{T.~Arisawa}
\affiliation{Waseda University, Tokyo 169, Japan}
\author{A.~Artikov}
\affiliation{Joint Institute for Nuclear Research, RU-141980 Dubna, Russia}
% ...

\collaboration{CDF Collaboration}
\altaffiliation[With visitors from]{
\ensuremath{^{a}}University of British Columbia, Vancouver, BC V6T 1Z1, Canada,
\ensuremath{^{b}}Istituto Nazionale di Fisica Nucleare, Sezione di Cagliari, 09042 Monserrato (Cagliari), Italy,
\ensuremath{^{c}}University of California Irvine, Irvine, CA 92697, USA,
\ensuremath{^{d}}Institute of Physics, Academy of Sciences of the Czech Republic, 182~21, Czech Republic,
\ensuremath{^{e}}CERN, CH-1211 Geneva, Switzerland,
\ensuremath{^{f}}Cornell University, Ithaca, NY 14853, USA,
\ensuremath{^{g}}University of Cyprus, Nicosia CY-1678, Cyprus,
\ensuremath{^{h}}Office of Science, U.S. Department of Energy, Washington, DC 20585, USA,
\ensuremath{^{i}}University College Dublin, Dublin 4, Ireland,
\ensuremath{^{j}}ETH, 8092 Z\"{u}rich, Switzerland,
\ensuremath{^{k}}University of Fukui, Fukui City, Fukui Prefecture, Japan 910-0017,
\ensuremath{^{l}}Universidad Iberoamericana, Lomas de Santa Fe, M\'{e}xico, C.P. 01219, Distrito Federal,
\ensuremath{^{m}}University of Iowa, Iowa City, IA 52242, USA,
\ensuremath{^{n}}Kinki University, Higashi-Osaka City, Japan 577-8502,
\ensuremath{^{o}}Kansas State University, Manhattan, KS 66506, USA,
\ensuremath{^{p}}Brookhaven National Laboratory, Upton, NY 11973, USA,
\ensuremath{^{q}}Queen Mary, University of London, London, E1 4NS, United Kingdom,
\ensuremath{^{r}}University of Melbourne, Victoria 3010, Australia,
\ensuremath{^{s}}Muons, Inc., Batavia, IL 60510, USA,
\ensuremath{^{t}}Nagasaki Institute of Applied Science, Nagasaki 851-0193, Japan,
\ensuremath{^{u}}National Research Nuclear University, Moscow 115409, Russia,
\ensuremath{^{v}}Northwestern University, Evanston, IL 60208, USA,
\ensuremath{^{w}}University of Notre Dame, Notre Dame, IN 46556, USA,
\ensuremath{^{x}}Universidad de Oviedo, E-33007 Oviedo, Spain,
\ensuremath{^{y}}CNRS-IN2P3, Paris, F-75205 France,
\ensuremath{^{z}}Universidad Tecnica Federico Santa Maria, 110v Valparaiso, Chile,
\ensuremath{^{aa}}The University of Jordan, Amman 11942, Jordan,
\ensuremath{^{bb}}Universite catholique de Louvain, 1348 Louvain-La-Neuve, Belgium,
\ensuremath{^{cc}}University of Z\"{u}rich, 8006 Z\"{u}rich, Switzerland,
\ensuremath{^{dd}}Massachusetts General Hospital, Boston, MA 02114 USA,
\ensuremath{^{ee}}Harvard Medical School, Boston, MA 02114 USA,
\ensuremath{^{ff}}Hampton University, Hampton, VA 23668, USA,
\ensuremath{^{gg}}Los Alamos National Laboratory, Los Alamos, NM 87544, USA,
\ensuremath{^{hh}}Universit\`{a} degli Studi di Napoli Federico I, I-80138 Napoli, Italy
}
\noaffiliation

\begin{abstract}
We report on mass and lifetime measurements of several ground state
charmed and  bottom baryons, 
using a data sample corresponding to 9.6 $\textrm{fb}^{-1}$ 
from $p\bar p$ collisions at
$\sqrt{s}=1.96$ TeV, and
recorded with the Collider Detector at Fermilab.
Baryon candidates are reconstructed from
data collected with an online event selection designed 
for the collection of long-lifetime heavy-flavor decay products
and a second event selection designed to collect 
$J/\psi \rightarrow \mu^+ \, \mu^-$ candidates.
First evidence for the process
$\Omega_b^- \rightarrow \Omega_c^0 \, \pi^-$ is presented 
with a significance of $3.3\sigma$.
We measure the following baryon masses:
\begin{eqnarray}
  M(\Xi_c^{0}) = 
2470.85\pm0.24(\textrm{stat})\pm0.55(\textrm{syst}) \, \textrm{MeV}/c^2, \nonumber \\
  M(\Xi_c^{+}) = 
2468.00\pm0.18(\textrm{stat})\pm0.51(\textrm{syst}) \, \textrm{MeV}/c^2, \nonumber \\
M(\Lambda_b) =   
5620.15\pm0.31(\textrm{stat})\pm0.47(\textrm{syst}) \, \textrm{MeV}/c^2, \nonumber \\
M(\Xi_b^-) = 
5793.4\pm1.8(\textrm{stat})\pm0.7(\textrm{syst}) \, \textrm{MeV}/c^2, 
\nonumber  \\
M(\Xi_b^0) = 
5788.7\pm4.3(\textrm{stat})\pm1.4(\textrm{syst}) \, \textrm{MeV}/c^2,   
\, \textrm{and} \nonumber \\
M(\Omega_b^-) = 
6047.5\pm3.8(\textrm{stat})\pm0.6(\textrm{syst}) \, \textrm{MeV}/c^2.  
\nonumber 
\end{eqnarray}
The isospin splitting of the $\Xi_b^{-,0}$ states is found to be
$M(\Xi_b^-)-M(\Xi_b^0)=4.7\pm4.7(\textrm{stat})\pm0.7(\textrm{syst})$ MeV/$c^2$.
The isospin splitting of the $\Xi_c^{0,+}$ states is found to be
$M(\Xi_c^0)-M(\Xi_c^+)$ =  
$2.85\pm0.30(\textrm{stat})\pm0.04(\textrm{syst})$ MeV/$c^2$.
The following lifetime measurements are made:
\begin{eqnarray}
\tau(\Lambda_b) = 
1.565\pm0.035(\textrm{stat})\pm0.020(\textrm{syst}) \, \textrm{ps}, \nonumber \\
\tau(\Xi_b^-) = 
1.32\pm0.14(\textrm{stat})\pm0.02(\textrm{syst}) \, \textrm{ps}, \nonumber \\
\tau(\Omega_b^-) =
1.66^{+0.53}_{-0.40}(\textrm{stat})\pm0.02(\textrm{syst}) \, \textrm{ps}. \nonumber 
\end{eqnarray}

\end{abstract}
\maketitle

\section{Introduction}

The quark model describes 
the spectroscopy of hadrons with great success.  
In particular, this has been the case for the $D$ and 
$B$ mesons, where all of the 
ground states have been observed \cite{PDG}.
The spectroscopy of $c$ baryons also agrees well with the quark model, 
and a  rich
spectrum of baryons containing $b$ quarks is predicted 
\cite{Jenkins,Karliner}.
The accumulation of large data sets from the Tevatron and 
Large Hadron Collider
has  made possible the observation and measurements of most of the
$b$-baryon ground states containing a single heavy 
quark 
\cite{D0_Xi_b, CDF_Xi_b,CDF_Sigma_b,D0_Omega_b,CDF_Omega_b,CDF_Xib0,LHCb_Xib}
and several resonant states
\cite{CDF_Sigma_b,CMS_Xib0,LHCb_LambdabStar,CDF_LambdabStar}.  
The samples
of most $b$ baryons accumulated to date are small, and the 
measurements of the properties
of these particles are limited by the sample size.  
The exception to this is the $\Lambda_b$ baryon, 
where the reconstructed samples
are now large enough to probe its properties with precision.
Early measurements of the $\Lambda_b$ lifetime disagreed with
predictions from heavy-quark expansion theory, if compared with the 
$B^0$ lifetime \cite{Gabbiani}.  
With the large samples that are now available, precision measurements 
of the $\Lambda_b$ lifetime  are providing a strong test
of the heavy-quark expansion 
in describing $b$ hadrons \cite{LHCb_Lambdab_life}. 

In this paper, we report the 
measurements of mass and lifetime, $\tau$, for the $\Lambda_b$, 
$\Xi^{-}_b$ and $\Omega^-_b$ baryons through the decay processes
$\Lambda_b \rightarrow J/\psi \, \Lambda$,
$\Xi^{-}_b \rightarrow J/\psi \, \Xi^-$, and
$\Omega^{-}_b \rightarrow J/\psi \, \Omega^-$.
Mass
measurements of the $\Xi_c^{0}$, $\Xi_c^{+}$, $\Xi^{-}_b$, and $\Xi^{0}_b$
are made by reconstructing the processes
$\Xi_c^{0} \rightarrow \Xi^- \, \pi^+$,
$\Xi_c^{+} \rightarrow \Xi^- \, \pi^+ \, \pi^+$,
$\Xi_b^{-} \rightarrow \Xi_c^0 \, \pi^-$, and
$\Xi_b^{0} \rightarrow \Xi_c^+ \, \pi^-$.
In addition, we report first evidence for the process
$\Omega_b^{-} \rightarrow \Omega_c^0 \, \pi^-$, 
$\Omega_c^{0} \rightarrow \Omega^- \, \pi^+$.
These decay chains are reconstructed using the processes
$J/\psi \rightarrow \mu^+ \, \mu^-$,
$\Xi^- \rightarrow \Lambda \, \pi^-$,
$\Omega^- \rightarrow \Lambda \, K^-$, and
$\Lambda \rightarrow p \, \pi^-$.
Charge conjugate modes are implicitly included.
These measurements are made in
$p\overline{p}$ collisions 
at a center-of-mass energy of 1.96 TeV using the
Collider Detector at Fermilab (CDF II), 
corresponding to an integrated luminosity of
9.6 fb$^{-1}$.  
This paper uses the full CDF II data set collected during the 2001-11 
operation 
of the Tevatron.  



The strategy of this analysis 
is to calibrate and check the measurement technique on the better known
$b$-meson states and then to extend the method to property measurements 
of the $b$ baryons reconstructed from the same data.
All mass and lifetime measurements are performed on the 
$B^+ \rightarrow J/\psi \, K^{+}$ and 
$B^0 \rightarrow J/\psi \, K^{*}(892)^{0}$, 
$K^{*}(892)^{0}\rightarrow K^+\pi^-$ 
final states to provide a large sample
for comparison to the world-average values.  
The decay mode $B^0 \rightarrow J/\psi \, K^{0}_S$, 
$K^{0}_S \rightarrow \pi^+ \, \pi^-$ is also used as a reference
process.  
Although its sample size is smaller than the samples of $B^+$ 
and other $B^0$
decays, this is an appropriate reference because
the $K^{0}_S$ is reconstructed  from charged particles that
are significantly displaced from the collision, similar to the
final-state particles from the $b$-baryon decays studied in this work.

We begin with a brief description of the detector and its simulation
in Sec.\ \ref{sect:Detector}.
In Sec.\ \ref{sect:Reconstruction}, the reconstruction
of $J/\psi$ mesons, neutral $K$ mesons, hyperons, and $b$ hadrons is described.
In Sec.\ \ref{sect:Properties}, we present 
measurements of the properties of the $\Xi_c^{0,+}$,
$\Lambda_b$, $\Xi^{-,0}_b$, and $\Omega^-_b$ baryons, which
include particle masses and lifetimes. 
We conclude in Sec.\ \ref{sect:Conclusions} with a summary of the results.

\section{Detector Description and Simulation\label{sect:Detector}}

The CDF II detector has been described in detail elsewhere 
\cite{CDF_detector}.  
This analysis primarily relies upon the charged-particle tracking and 
muon-identification systems.  
The tracking system consists of 
four different detector subsystems that operate inside a 1.4 T solenoid
with its axis parallel to the beamline.
The first of these is a single layer of silicon strip detectors (L00) 
at a radius of $1.35 - 1.6$~cm from the axis of the solenoid. 
It measures charged-particle positions (hits) in the 
transverse 
view with respect to the beam, 
which is parallel to the $z$ direction.
A five-layer silicon detector (SVX\ II).  
 surrounding L00
measures hits at
radii of 2.5 to 10.6~cm \cite{CDF_Silicon}. 
Each of these layers
provides a transverse measurement
and a stereo measurement of $90^{\circ}$
(three layers) or  $\pm 1.2^{\circ}$ (two layers)
with respect to the beam direction.
The outermost silicon detector lies between 19 cm and 30 cm radially, and
provides one or two hits, 
depending on the track pseudorapidity ($\eta$), where  
$\eta \equiv -\ln [ \tan( \theta /2) ]$, with $\theta$ 
being the angle between the particle
momentum and the proton-beam direction.
An open-cell drift chamber (COT)
completes the tracking system, and covers the radial region from
43 cm to 132 cm \cite{CDF_COT}.
The COT consists of 96 sense-wire layers, arranged in 8 superlayers 
of 12 wires each.  Four of these superlayers provide axial measurements and four
provide  stereo views of $\pm2^{\circ}$.  Transverse momentum, $p_T$, 
(defined as the component of the particle
momentum perpendicular to the proton-beam direction) of 
charged particles is measured in the COT with  a resolution of
$\sigma(p_{T})/p_{T}^2 = 0.0017$ [GeV/$c$]$^{-1}$.

Electromagnetic and hadronic calorimeters surround the solenoid coil.
 Muon candidates from the decay $J/\psi \rightarrow \mu^+ \mu^-$
 are  identified by  two sets of drift  chambers located radially
 outside  the     calorimeters.
The central muon chambers
cover the region $ |\eta| <  0.6$, 
and  detect muons with 
$p_T > 1.4 $ GeV/$c$ \cite{CDF_Muon}.
A second muon system covers the region $  0.6 < |\eta| <  1.0$ and
detects  muons with $p_T > 2.0 $ GeV/$c$.  
Muon selection
is based on matching the measurements
from these chambers to COT tracks, both in
projected position and angle.

The analysis presented here  is based on events recorded with two
different online event-selection (trigger) algorithms.
The first is dedicated to the collection of a 
$J/\psi \rightarrow \mu^+ \mu^-$ sample.
The first level of the three-level trigger system requires two
muon candidates with matching tracks in the COT and muon chambers.
The second level imposes the requirement that muon candidates
have opposite electric charge 
and limits the accepted range of azimuthal opening angle.  The highest level 
of the $J/\psi$  trigger reconstructs the muon pair in software, and 
requires the invariant mass of the pair to fall within the range
$2.7-4.0$~GeV/$c^2$.

The second data set used is triggered by a system designed to collect
particle candidates 
that decay with lifetimes characteristic of heavy flavor hadrons.
The first level of this trigger requires two charged particles
in the COT with $p_T > 2.0 $ GeV/$c$.
In the second level of the trigger, the silicon vertex trigger 
\cite{SVT} is used to associate SVX II data with the tracks 
found in the COT to precisely measure the impact parameter 
(defined as the distance of closest approach in the transverse view)
with respect to the beamline.  The impact parameter
resolution (typically 40 $\mu$m) for
these tracks allows the isolation of a track sample that does
not originate directly from the $p\bar p$ collision \cite{CDF_Silicon}.  
The silicon vertex trigger requires two tracks with  
impact parameters $d$ in the range $0.1-1.0$~mm with 
respect to the beam and a point of intersection  at least 200 $\mu$m 
from  the beamline in the transverse view.
These and other requirements bias the trigger efficiency
toward candidates that have a long decay time.  Lifetime measurements
made with these data therefore require a careful study of these biases
and appropriate corrections.
The additional statistical power from lifetime measurements 
that use these data is insufficient to overcome the systematic 
uncertainty due to the trigger conditions.
Therefore, only mass measurements are extracted 
from the hadronic trigger data in this work.

The mass resolution and acceptance for the
$b$ hadrons used in this analysis are
studied with a Monte Carlo simulation that generates $b$ hadrons
consistent with CDF measurements of $p_T$ and rapidity distributions.
The final-state decay processes are simulated with 
the {\sc EvtGen} \cite{EvtGen} program, 
and all simulated $b$ hadrons are produced
without polarization.
The generated events are 
input to the detector and trigger simulation based on a 
{\sc GEANT3} description \cite{Geant}  
and processed through 
the same reconstruction and
analysis algorithms that are used for the data.

\section{Particle Reconstruction Methods \label{sect:Reconstruction}}

\subsection{$J/\psi$ reconstruction}
The analysis of the data 
obtained with the muon trigger
begins with a selection of well-measured
$J/\psi \rightarrow \mu^+ \mu^-$ candidates.
The trigger requirements are confirmed by selecting
 events that contain two
 oppositely  charged  muon candidates, each with matching 
 COT and muon-chamber tracks.
Both muon tracks are required to have associated position
measurements in at least three layers of the SVX\ II.
This data sample provides approximately $6.5\times 10^7$  $J/\psi$
candidates, measured with an average mass resolution of 
approximately $20$~MeV/$c^2$.
These candidates are required to have
a two-track invariant  mass within the range listed in
Table \ref{table:mass_range}.
 
\subsection{Neutral Hadron Reconstruction \label{sect:neutral_hadron}}

The reconstruction of $K^0_S$, $K^{*}(892)^{0}$, and $\Lambda$
candidates uses all particles with $p_T \, >$ 0.4~GeV/$c$
found in the COT that are not associated with muons used in the $J/\psi$ 
reconstruction or tracks used by the hadronic trigger.  
Pairs of oppositely charged particles are combined to identify these
neutral decay candidates.
Silicon detector information
is not used on these to avoid decay-length-dependent biases
on the reconstruction efficiency due to 
the long lifetimes of the particles. 
Candidate selection for these neutral states is 
based upon the mass calculated for each track pair,
which is required to fall within the ranges given 
in Table\ \ref{table:mass_range}
after the appropriate mass assignment is made for each track.
Backgrounds to the $K^0_S$~($c\tau \approx 2.7$~cm) and  
$\Lambda$~($c\tau \approx 7.9$~cm) signals \cite{PDG} 
are reduced by imposing requirements
on the transverse flight-distance, given
for neutral particles as
$f(h) \equiv (\vec {r}_{d} - \vec {r}_{o})
 \cdot \vec{p}_{T}(h)/|\vec{p}_{T}(h)|$, where 
$\vec{p}_{T}(h)$ is the transverse momentum of the hadron candidate, 
and $\vec {r}_{d}, \vec {r}_{o}$ are the transverse positions of the
decay point and point of origin, respectively.
The transverse flight-distance
of the $K^0_S$ and $\Lambda$ candidates with respect to the primary vertex
(defined as the beam position in the transverse view) is required to be
greater than 1.0 cm.

\begin{table}[ht]
\begin{center}
\caption{Mass ranges around the known mass values \cite{PDG}
used for the $b$-hadron decay products.
\label{table:mass_range}}
\begin{tabular}{cc}
\hline \hline
Resonance (final state) & Mass range (MeV/$c^2$)\\
\hline
 $J/\psi\, (\mu^+ \mu^-)$         &  $\pm80$ \\
 $K^{*}(892)^{0} \, (K^+ \pi^-)$  &  $\pm30$ \\
 $K^0_S \, (\pi^+ \pi^-)$         &  $\pm20$ \\
 $\Lambda \, (p \pi^-)$           &  $\pm9$ \\
 $\Xi^- \, (\Lambda \pi^-)$       &  $\pm9$ \\
 $\Omega^- \, (\Lambda K^-)$      &  $\pm8$ \\
 $\Xi_c^0 \, (\Xi^- \pi^+)$       &  $\pm30$ \\
 $\Xi_c^+ \, (\Xi^- \pi^+ \pi^+)$ &  $\pm25$ \\
 $\Omega_c^0 \, (\Omega^- \pi^+)$ &  $\pm30$ \\
\hline \hline
\end{tabular}
\end{center}
\end{table}

\subsection{Charged Hyperon Reconstruction}

For events that contain a $\Lambda$ candidate,
the remaining particles reconstructed in the COT, again without additional
silicon information,
are assigned the pion or kaon mass, and
$\Lambda \, \pi^-$ or $\Lambda \, K^-$ combinations 
are identified that are consistent with the decay process
$\Xi^- \rightarrow \Lambda \, \pi^-$ or 
$\Omega^- \rightarrow \Lambda \, K^-$.
Candidates are required to have a mass that is consistent 
with the ranges listed in 
Table \ref{table:mass_range}.  
Charged particles with $p_T$ 
as low as 0.4~GeV/$c$ are used for $\Xi^-$ reconstruction.
However, event simulation
indicates that the $p_T$ distribution of $K^-$ mesons produced
from $\Omega^-$ decays has a higher average value, and declines 
 more slowly, than the $p_T$ distribution of the
pions from $\Lambda$ or $\Xi^-$ decays.
Therefore, $p_T(K^-) > 1.0$ GeV/$c$ is required
for the $\Omega^-$ sample.

Several features of the track topology are used to reduce the
$\Xi^-$ and $\Omega^-$ backgrounds.
In order to improve the  
mass resolution for $\Xi^-$ and $\Omega^-$ candidates, 
the reconstruction requires a good  
fit of the three tracks that simultaneously 
constrains the $\Lambda$
decay products to the $\Lambda$ mass, and the
$\Lambda$ trajectory to intersect with the helix of 
the $\pi^-(K^-)$ originating from the $\Xi^-(\Omega^-)$ candidate.
In addition,
the transverse flight-distance of the $\Lambda$ candidate with respect to the
reconstructed decay vertex of the $\Xi^-(\Omega^-)$ candidate
is required to exceed 1.0 cm.
Due to the long lifetime of the $\Xi^-$ ($c\tau \approx$ 4.9 cm)
and $\Omega^-$ ($c\tau \approx$ 2.5 cm) particles \cite{PDG}, 
a transverse flight-distance
of at least 1.0 cm (corresponding to a measurement uncertainty of
approximately one standard deviation for a typical candidate) 
with respect to the primary vertex
is required.  
Transverse flight-distance for charged particles is defined as
the arc length from the point of closest approach to the origin
to the decay point.
Possible kinematic reflections are removed from the $\Omega^-$ sample
by requiring that the combinations in the sample fall outside the 
$\Xi^-$ mass range listed in Table\ \ref{table:mass_range} when the candidate
$K^-$ track is assigned the mass of the $\pi^-$.  
In instances where the correct vertex assignment for the decay tracks 
is ambiguous, 
a fit is performed for all configurations and
a single, preferred candidate is chosen by retaining only
the fit combination with the lowest $\chi^2$.

\subsection{Charmed Hyperon Reconstruction}
The $\Xi^-$ and $\Omega^-$ candidates are used to reconstruct 
the processes
$\Xi_c^0 \rightarrow \Xi^- \, \pi^+$,
$\Xi_c^+ \rightarrow \Xi^- \, \pi^+ \, \pi^+$, and
$\Omega_c^0 \rightarrow \Omega^- \, \pi^+$.  
Each $c$-baryon candidate is subjected to a simultaneous fit of all the
tracks in the decay process that constrains the track intersections
and decay product momenta
 to be consistent with the appropriate decay topology.
In addition, the tracks from the $\Lambda$ decay are constrained 
to the known $\Lambda$ mass.

The kinematic properties of $\Xi^-$ and $\Omega^-$ 
decays and the lower $p_T$ limit of 0.4~GeV/$c$
on the final-state tracks cause the majority of accepted 
charged hyperon 
candidates to have $p_T >$~1.5~GeV/$c$.  
This fact, along with the long lifetimes of the $\Xi^-$ 
and $\Omega^-$,
results in a significant fraction of hyperon candidates having
decay vertices located several centimeters radially outward from the 
beam position.
Therefore, we 
refine the charged-hyperon  reconstruction by using the 
improved determination of its trajectory available from tracking 
these particles in the silicon detector.  
The $\Xi^-(\Omega^-)$ point of origin, point of decay, and momentum
 obtained from the
full four- or five-track fit are used to define a helix that
serves as the seed for an algorithm that associates silicon detector
hits with the charged-hyperon track.  
Charged-hyperon
candidates with track measurements in at least one layer of the 
silicon detector
have excellent impact parameter resolution (average of 60 $\mu$m) for the
charged hyperon track. 

\begin{figure}[hbt]
\psfig{figure=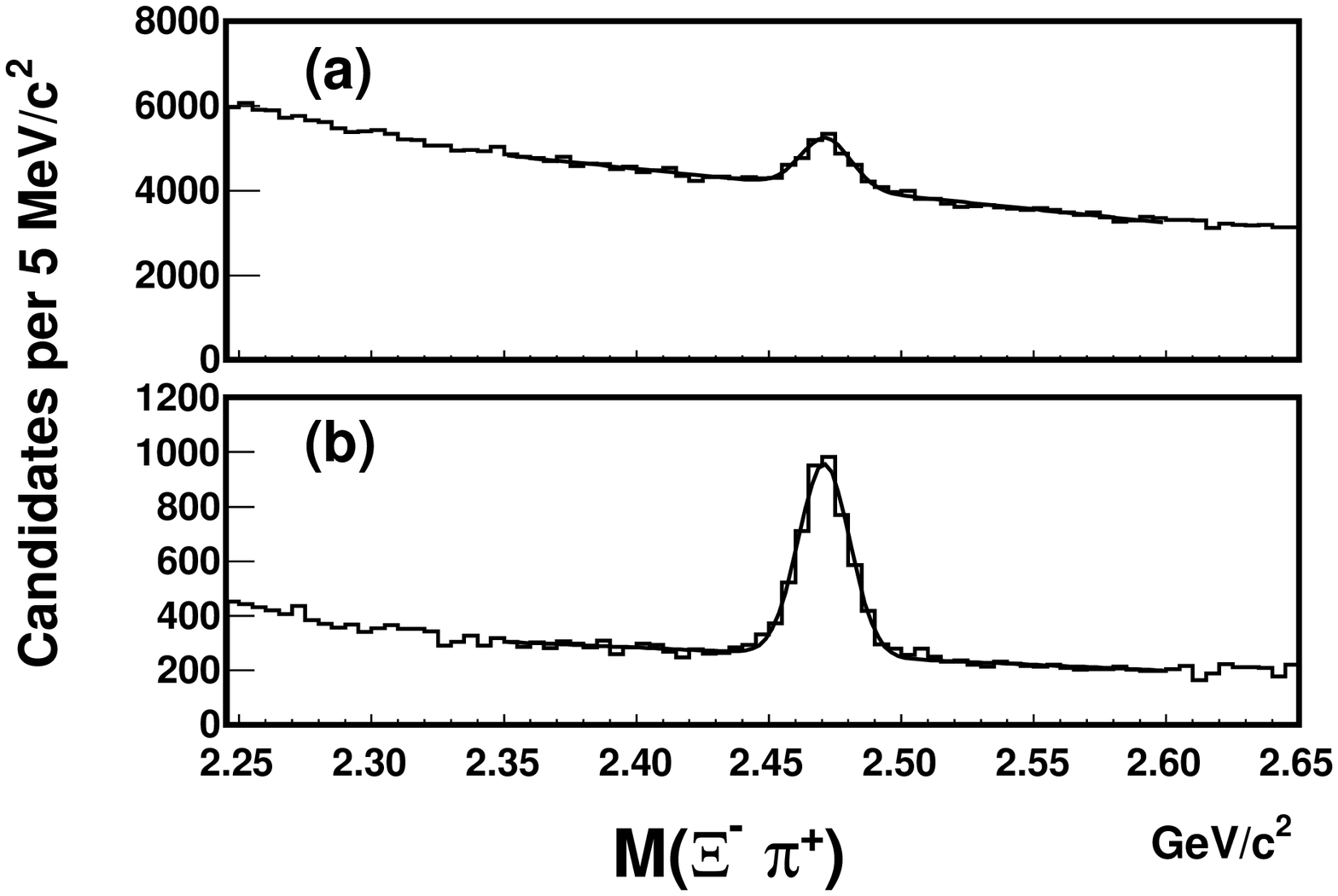,height=2.25in} 
\psfig{figure=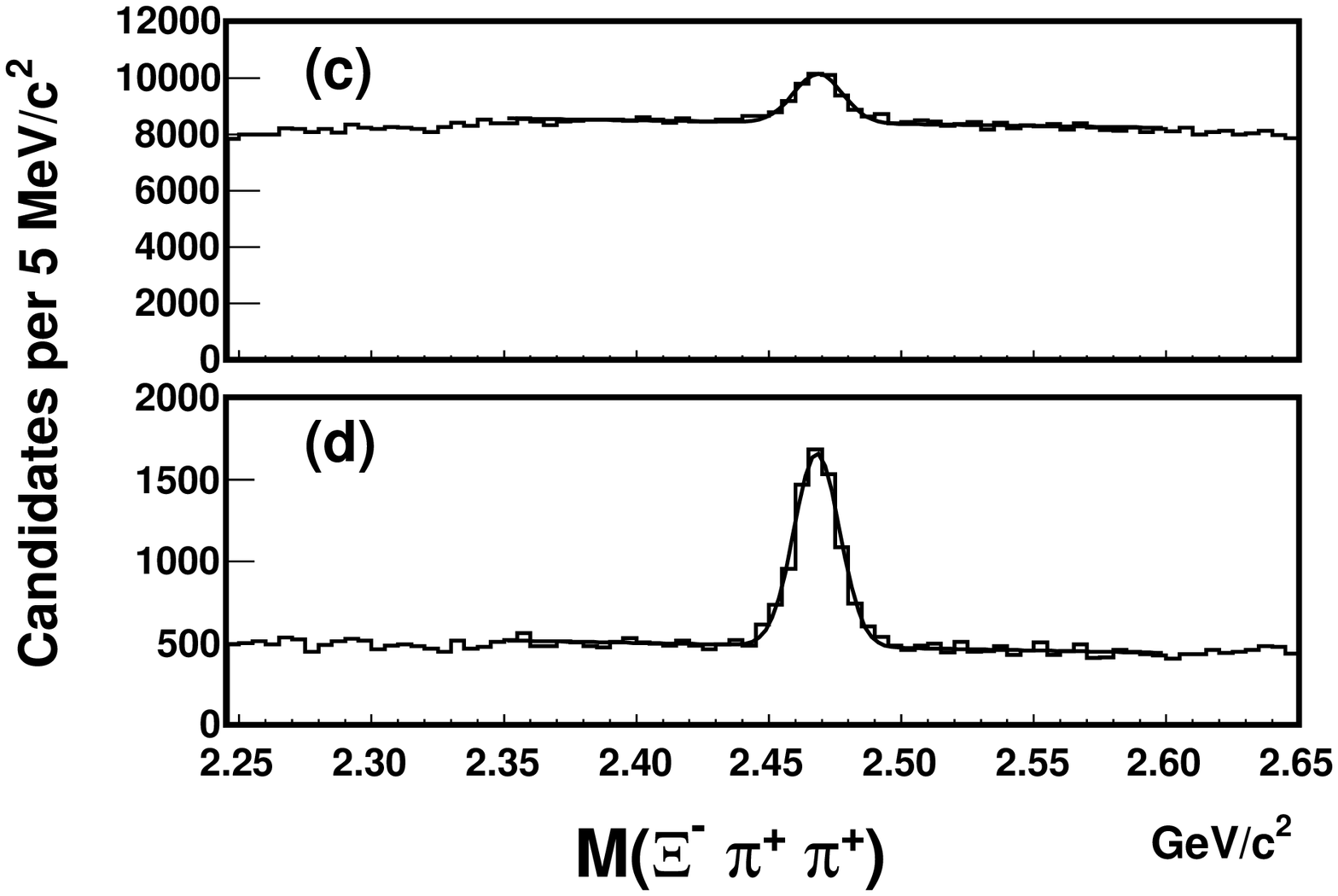,height=2.25in} 
\psfig{figure=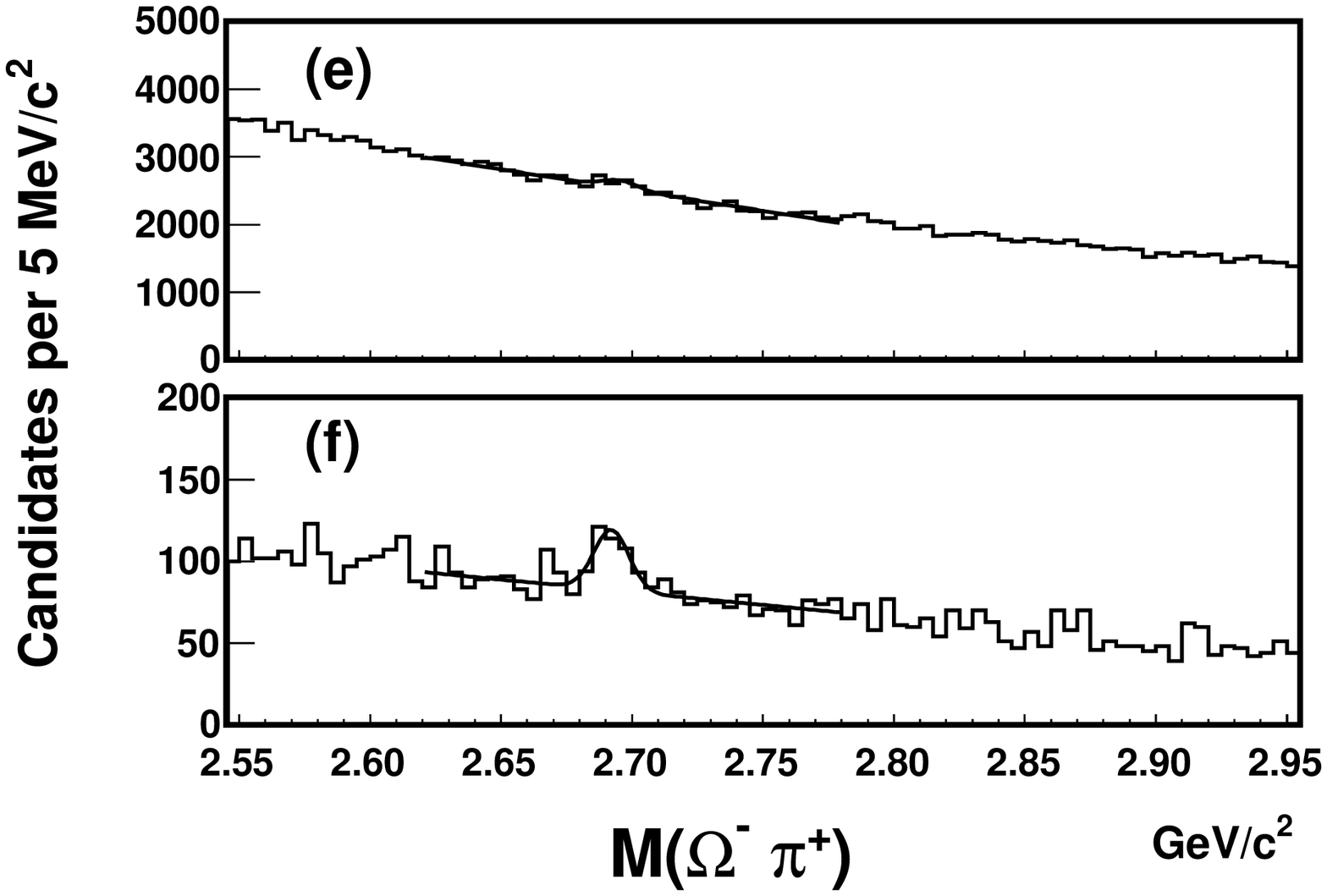,height=2.25in} 
\caption{Distribution of $\Xi^- \, \pi^+$
mass for (a) all candidates and (b) the subset 
where the $\Xi^-$ is tracked in the silicon detector and the impact
parameter with 
respect to the beamline is less than 100 $\mu$m. 
Panels (c,d) and (e,f) show similar distributions for
 $\Xi^- \, \pi^+\, \pi^+$ and $\Omega^- \, \pi^+$ 
candidates, respectively.
 \label{fig:fig_1}}
\end{figure}

Mass distributions are shown in Fig.\ \ref{fig:fig_1}
for all combinations and for the subset where the $\Xi^-$ or $\Omega^-$
track reconstruction is improved by using at least one hit in the SVX II
and the impact parameter
of the $c$ baryon with respect to the beam is less than 100 $\mu$m.  
The improvement in charmed-hyperon purity is evident.
An estimate of the yield in each case is made by performing a binned fit
on these distributions, which 
models the data with a Gaussian signal and linear
background.  Due to the small sample size, the 
Gaussian width term for the $\Omega_c^0$ is fixed at 8 MeV/$c^2$, 
which is the resolution predicted by the event simulation.
The background under each signal is estimated by integrating the
background function from the fit over the range $\pm2\sigma$ around 
the signal mass, where $\sigma$ is the characteristic width of the Gaussian
signal.  Signal yields and purity, defined as signal-to-background ratio, 
are listed in Table \ref{table:charm_purity} for all fits.
\begin{table}[ht]
\begin{center}
\caption{Signal yields and purity for charmed hyperon samples.
Only statistical uncertainties are listed.
\label{table:charm_purity}}
\begin{tabular}{ccccc}
\hline \hline
State & \multicolumn{2}{c}{Full sample} & 
\multicolumn{2}{c}{Tracked in SVX II} \\
\hline 
 & Yield & Purity & Yield & Purity \\
$\Xi_c^0$    & $5614\pm247$ & $0.15\pm0.01$ & $3412\pm 84$ & $0.63\pm0.01$ \\
$\Xi_c^+$    & $7984\pm354$ & $0.11\pm0.01$ & $5065\pm104$ & $0.61\pm0.01$ \\
$\Omega_c^0$ &  $416\pm135$ & $0.03\pm0.01$ &  $124\pm 31$ & $0.22\pm0.05$ \\
\hline \hline
\end{tabular}
\end{center}
\end{table}
A requirement of silicon-detector information
on the $\Xi^-$ candidate track is
approximately 60\% efficient for the inclusive $\Xi_c$ baryon sample.
Signal purity increases markedly when the silicon is used.
The efficiency of this requirement on the $\Omega^-$ signal is lower,
as expected from the shorter lifetime of the $\Omega^-$.
A substantial fraction of $\Omega^-$ decay prior to 
reaching the SVX II, so they fall outside the acceptance of the
detector.  However, the use of silicon when it is available provides
a significant improvement in signal purity.
Consequently, 
charmed hyperon candidates retained for further analysis
are required to have a $\Xi^-$ or $\Omega^-$ candidate
measured in the silicon detector. At least one $\pi^+$ with 
$p_T > 2.0$ GeV/$c$ and $d \,> \,100 \, \mu$m is required, for consistency 
with the trigger.  
We also require $p_T > 4.0$~GeV/$c$ and
$ct > 100 \, \mu$m for the charmed hyperon candidates, 
where $t$ is the measured decay time given by $t = f \, M/p_T$,
$M$ is the reconstructed mass of the candidate, 
and $f$ is the transverse flight distance defined in 
Sec.~\ref{sect:neutral_hadron}.

\subsection{$b$ baryon reconstruction}

A good fit is required on the final-state tracks of all $b$-baryon candidates 
that constrains them to originate from the vertices appropriate for 
the particular decay channel being considered.  
In addition,
we require $ct > 100 \, \mu$m and $d \,< \, 100 \, \mu$m
for each $b$-baryon candidate,
to remove prompt and poorly reconstructed candidates.  All hadron
decay products used in the $b$-baryon reconstruction are required
to have a measured mass consistent with the known values,
according to the ranges listed in 
Table \ref{table:mass_range}.

Several selection criteria are used that are common
to all $b$-hadron candidates with a $J/\psi$ meson
in the final state. 
We require the transverse momentum of the $b$ hadron to exceed
6.0~GeV/$c$ and $p_T(h) > 2.0 $~GeV/$c$,
where $h$ is the hadron accompanying the $J/\psi$ meson.
These requirements reduce combinatorial background.  
In addition, the final-state fit  constrains  
the mass of the $\mu^+ \, \mu^-$ pair to the 
known mass of the $J/\psi$ meson \cite{PDG}.  
The hadron tracks are reconstructed without 
silicon-detector information.  Therefore, all 
$b$-hadron decay position information is derived solely
from the muons, and the decay-time resolution is simular for all
$b$ hadrons in this data set.

The hadronic trigger data provides a sample of $\Xi_b$ baryons through
the decay channel $\Xi_b \rightarrow \Xi_c \, \pi^-$, 
$\Xi_c \rightarrow \Xi^- \, \pi^+ \, (\pi^+)$, 
$\Xi^- \rightarrow \Lambda \, \pi^-$,
and $\Lambda \rightarrow p \, \pi^-$.
A similar decay chain is used for $\Omega^-_b$ reconstruction.
The final-state track fit used in these decay processes
includes a constraint on the
$\Lambda$ decay products to the known $\Lambda$ mass.
The $\pi^-$ candidates from the $b$-baryon decay are required to 
have electric charge opposite to the $\Lambda$ baryon number, and
to be consistent with having satisfied the trigger 
by having $p_T > 2.0$ GeV/$c$ and $d \, > \, 100 \, \mu$m.
The backgrounds under the $\Xi_b$ states are also reduced by restricting the
sample based on the measured decay time of the $\Xi_c$ candidates to the range
$-2\sigma_{t} \, < \, t(\Xi_c) \, < \, 3 \tau_0(\Xi_c) + 2\sigma_{t}$, 
where $\sigma_{t}$ is the calculated uncertainty on the decay time and
$\tau_0(\Xi_c)$ is the known lifetime of the appropriate $\Xi_c$ 
baryon \cite{PDG}.

\begin{figure}[hbt]
\psfig{figure=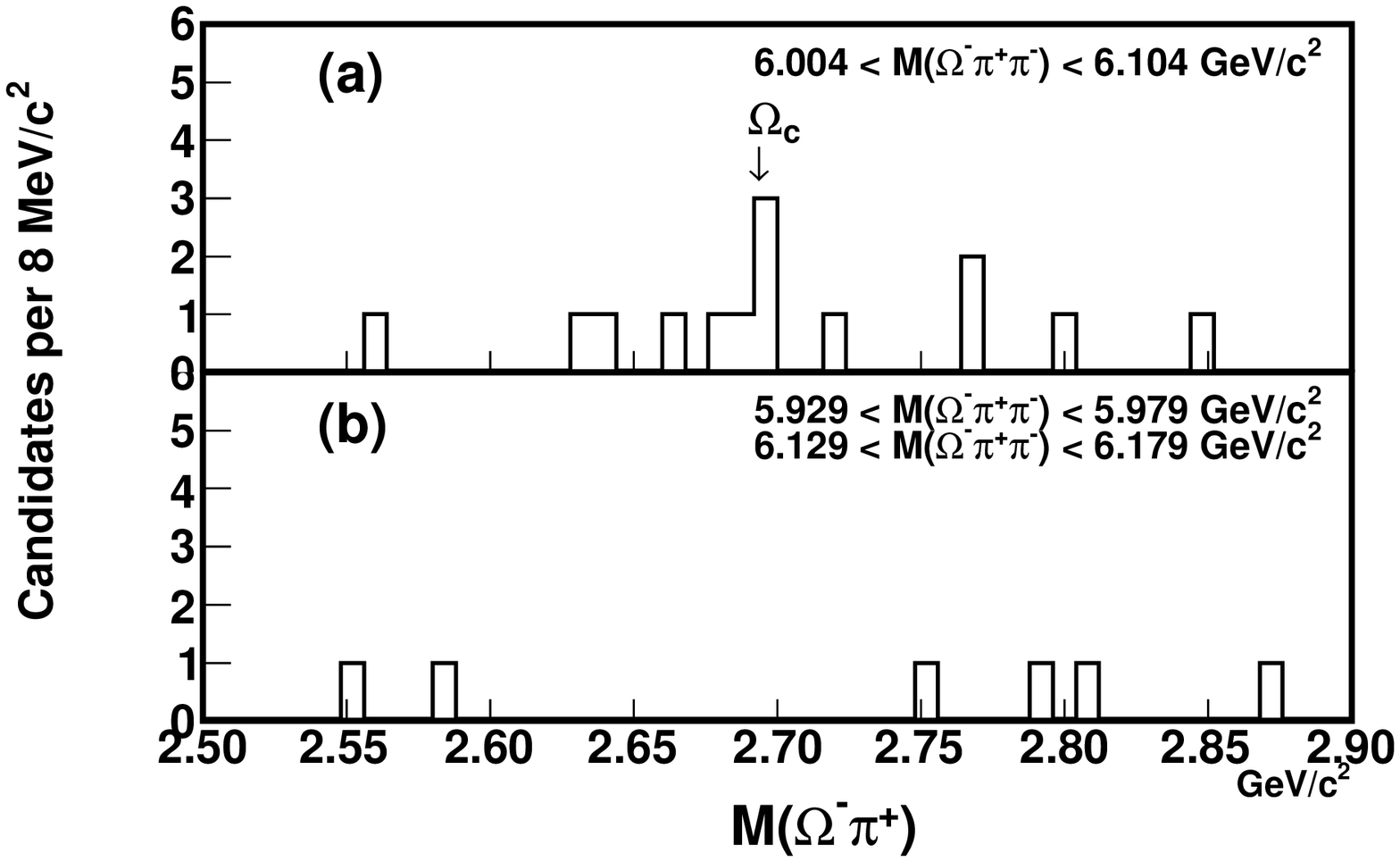,height=2.25in} 
\psfig{figure=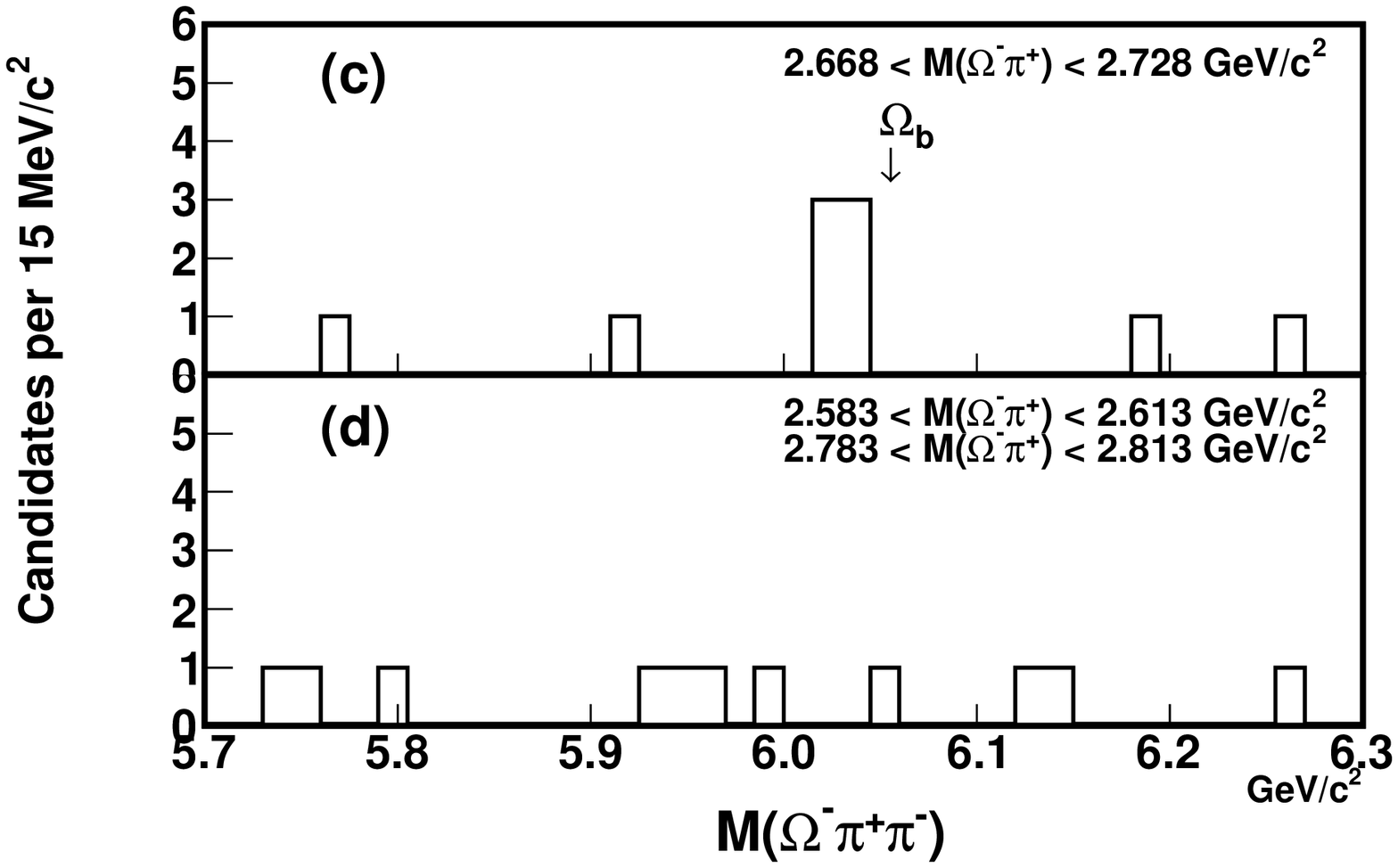,height=2.25in} 
\caption{Distribution of $\Omega^- \, \pi^+$ (a,b) and 
$\Omega^- \, \pi^+ \, \pi^-$ 
(c,d) mass for candidates obtained from the $\Omega_b^-$ selection. 
Panel (a) shows the $\Omega^- \, \pi^+$ mass for candidates consistent with 
the $\Omega_b^- \rightarrow \Omega^- \pi^+ \, \pi^-$ signal region;
panel (b) shows the $\Omega^- \, \pi^+$ mass for candidates restricted to the
 $\Omega_b^-$ mass sidebands.
Panel (c) shows the $\Omega^- \, \pi^+ \, \pi^-$ mass for candidates 
consistent with the $\Omega_c^0 \rightarrow \Omega^- \, \pi^+ $ signal region;
panel (d) shows the $\Omega^- \, \pi^+ \, \pi^-$ mass for candidates 
restricted to the $\Omega_c^0$ mass sidebands.
 \label{fig:fig_2}}
\end{figure}

The lifetime of the $\Omega_c^0$ is so short 
($c\tau \approx 21 \, \mu$m) \cite{PDG}
that the tracking system has no ability to resolve it.  Consequently,
no $t(\Omega_c^0)$ requirement is made in the selection of
$\Omega_c^0 \, \pi^-$ combinations.  Figure \ref{fig:fig_2} shows 
$\Omega^- \, \pi^+$ and $\Omega^- \, \pi^+\, \pi^-$ mass distributions
of selected combinations.  Figure \ref{fig:fig_2}(a) shows 
the distribution of all $\Omega^- \, \pi^+$ combinations that,
combined with a $\pi^-$ candidate, yield a mass 
within 50 MeV/$c^2$ of the
$\Omega_b^-$ mass previously measured by this experiment \cite{CDF_Omega_b}.
The known mass of the $\Omega_c^0$ \cite{PDG} is also indicated.
The $\Omega^- \, \pi^+$ combinations shown in 
Fig.\ \ref{fig:fig_2}(b) are chosen from two 
$\Omega^- \, \pi^+\, \pi^-$ mass sidebands, selected to be 
50 MeV/$c^2$ in width and centered
at $\pm100$~MeV/$c^2$ from the $\Omega_b^-$ mass.  
There is a clear indication of $\Omega_c^0$ candidates in events
where the  $\Omega^- \, \pi^+\, \pi^-$ mass is consistent with 
the $\Omega_b^-$ mass,
whereas no enhancement compatible with an $\Omega_c^0$ signal
appears in the background sample.  
A similar comparison
is made between Figs. \ref{fig:fig_2}(c) and \ref{fig:fig_2}(d),
where the $\Omega^- \, \pi^+\, \pi^-$ mass is shown for candidates
consistent with $\Omega_c^0$ decays and candidates from the 
sidebands of the $\Omega^- \, \pi^+$ mass distribution.

\subsection{Evidence for the 
$\Omega_b^- \rightarrow \Omega_c^0 \, \pi^-$ decay}

The indication of an $\Omega_b^-$ signal in the $\Omega_c^0 \, \pi^-$
mass distribution
shown in Fig.\ \ref{fig:fig_2}(c) requires additional consideration.  
Because the process $\Omega_b^- \rightarrow \Omega_c^0 \, \pi^-$
has never been observed, a standard significance test is performed where
the mass distribution is fit once with  a signal amplitude
that is allowed to float
and once where it is fixed to zero (the null hypothesis).
The signal mass used is fixed and the measurement 
resolution is fixed to 20~MeV/$c^2$, as determined by the simulation.
Two different $\Omega_b^-$ mass assumptions are used,
corresponding to the value measured in this work,
and the value recently measured by the LHCb Collaboration 
\cite{LHCb_Xib}, in order to assess the sensitivity of the significance
to the mass value.
Twice the change in the logarithm of the fit likelihood between 
the null and 
floating signal hypotheses, 
$2\Delta \ln {\cal L}$, is found to 
be $10.3$ and $13.3$ for the different $\Omega_b^-$ mass assumptions.  

The probability that the signal shown in Fig.\ \ref{fig:fig_2}(c) arises 
from a background fluctuation is obtained from a simple simulation of
the distribution of ten independent mass values generated uniformly over the
range used in Fig.\ \ref{fig:fig_2}(c).  The generated unbinned 
distribution is then 
fit with the likelihood function twice, as is done with the data.  
The value of $2\Delta \ln {\cal L}$
between the two fits is then recorded.  The process is repeated
$10^7$ times and values of $2\Delta \ln {\cal L} = 10.3$ or greater
occurs with a frequency of $5.5~\times~10^{-4}$.  This corresponds to 
a single sided fluctuation of a Gaussian distribution of $3.3 \sigma$,
corresponding to evidence for
the process $\Omega_b^- \rightarrow \Omega_c^0 \, \pi^-$.

\section{Particle Properties \label{sect:Properties}}
The mass and lifetime of the $b$ hadrons are measured by a fit
with data binned in decay time, but
not in mass \cite{CDF_Omega_b}.  
The mass and signal yield in each $ct$ bin are found by maximizing
a likelihood given by
\begin{equation}
{\cal L}  =  \prod_j^{N_{b}} \prod_i^{N_j} \left[ f_j{\cal P}^s_i + 
(1-f_j) {\cal P}^b_{i,j} \right],
\label{eq:likely}
\end{equation}
where $N_{b}$ is the number of $ct$ bins chosen for the fit,
$N_j$ and $f_j$ are the numbers of candidates 
and the signal fraction, respectively, for time bin $j$, and
${\cal P}^s_i$ and ${\cal P}^b_{i,j}$ are the mass
probability density functions
for the signal and background, respectively, for candidate $i$.
The signal probability distribution is given by
\begin{equation}
{\cal P}^s_i = (1 - \alpha)G(m_i,m_0,s_0\sigma^m_i) + 
\alpha G(m_i,m_0,s_1\sigma^m_i),
\end{equation}
where $G$ are Gaussians with average $m_0$; $m_i$ and $\sigma^m_i$ are
the measured mass and uncertainty for candidate $i$; 
and $\alpha, s_{0}$, and
$s_1$ are parameters determined in the fit 
that describe, respectively, the relative contribution from each Gaussian 
and possible deviations between the calculated and true mass uncertainty.
The background is modeled as
\begin{equation}
{\cal P}^b_{i,j} = \sum_{n=0}^{2} a_{n,j} P_n(m_i),
\end{equation}
where $P_n(m_i)$ are orthonormal polynomials of order $n$, which are
normalized over the range of the fit, and $a_{n,j}$ 
are constants obtained in the fit.  The background constants are
obtained independently for each time range $j$.
The overall 
normalization is assured by fixing $a_{0,j} = 1 - \sum_{n=1}^{2}a_{n,j}$.

The lifetime is determined by virtue of the fact that the fractional 
occupancy of each particular range of $ct$ implies a specific lifetime
for a particular measurement resolution.  This is implemented in 
a two-step process, which begins by maximizing the likelihood function
in the mass distributions given in Eq. (\ref{eq:likely}).  
In the second step,
all parameters obtained in the mass fits are fixed and
an additional lifetime term is added to the likelihood, rewriting it as
\begin{equation}
{\cal L}  =  \prod_j^{N_{b}} G(R_j,w_j(\tau,\sigma_{\tau}),\sigma^R_j)
\prod_i^{N_j} \left[f_j{\cal P}^s_i + (1-f_j) {\cal P}^b_{i,j} \right], \\
\end{equation}
where $R_j = f_j N_j/\sum_j f_j N_j$, $\sigma^R_j$ is the uncertainty on $R_j$,
and $w_j(\tau,\sigma_{t})$ is the predicted fractional occupancy
in each time range for lifetime $\tau$ measured with uncertainty
$\sigma_{t}$.  The predicted occupancy is found by integrating
the decay-time distribution convoluted with the 
decay-time-measurement resolution, 
which is assumed to be Gaussian and is calculated analytically.  
Decay-time bins are chosen
to have approximately equal occupancy for the initial lifetime
 chosen for the fit.  
The highest bin has no upper bound. 

There are several advantages to this technique over the usual method of 
simultaneously fitting the signal and background lifetimes.  
The only distribution where signal and background components are fit 
together is the mass distribution in which these components
are clearly discriminated given the
differences in shape between the narrow signal distributions and 
the smooth, quasi-uniform background distributions.
The number
of parameters in the fit is limited, typically two for the mass, 
one for the lifetime, and two for each decay time bin to 
account for the yield and
slope of the background.  Finally, this method does not require a model of the 
effective decay-time distribution of the background.  
 
Unless otherwise mentioned, 
all the implementations of the fit used in this analysis set the lowest
limit of the lowest decay time bin to $ct = 100 \, \mu$m,
use an initial lifetime estimate of
$ct = 450 \, \mu$m,
and utilize four decay-time bins ($N_b = 4$).
Also, the decay time resolution is set so $\sigma_{ct} = 30 \, \mu$m, 
which is typical of all candidates and is discussed further in 
Sec \ref{sect:systematics}.  The lower limit of all mass fit
ranges is chosen to be approximately one pion mass lower
than the expected average mass of the reconstructed hadron in order
to avoid backgrounds due to partially reconstructed states.
Upper limits are 
chosen to obtain a reasonable estimate of the background.

\subsection{The $B$ meson reference signals \label{sect:B_refer}}
\begin{figure}[hbt]
\psfig{figure=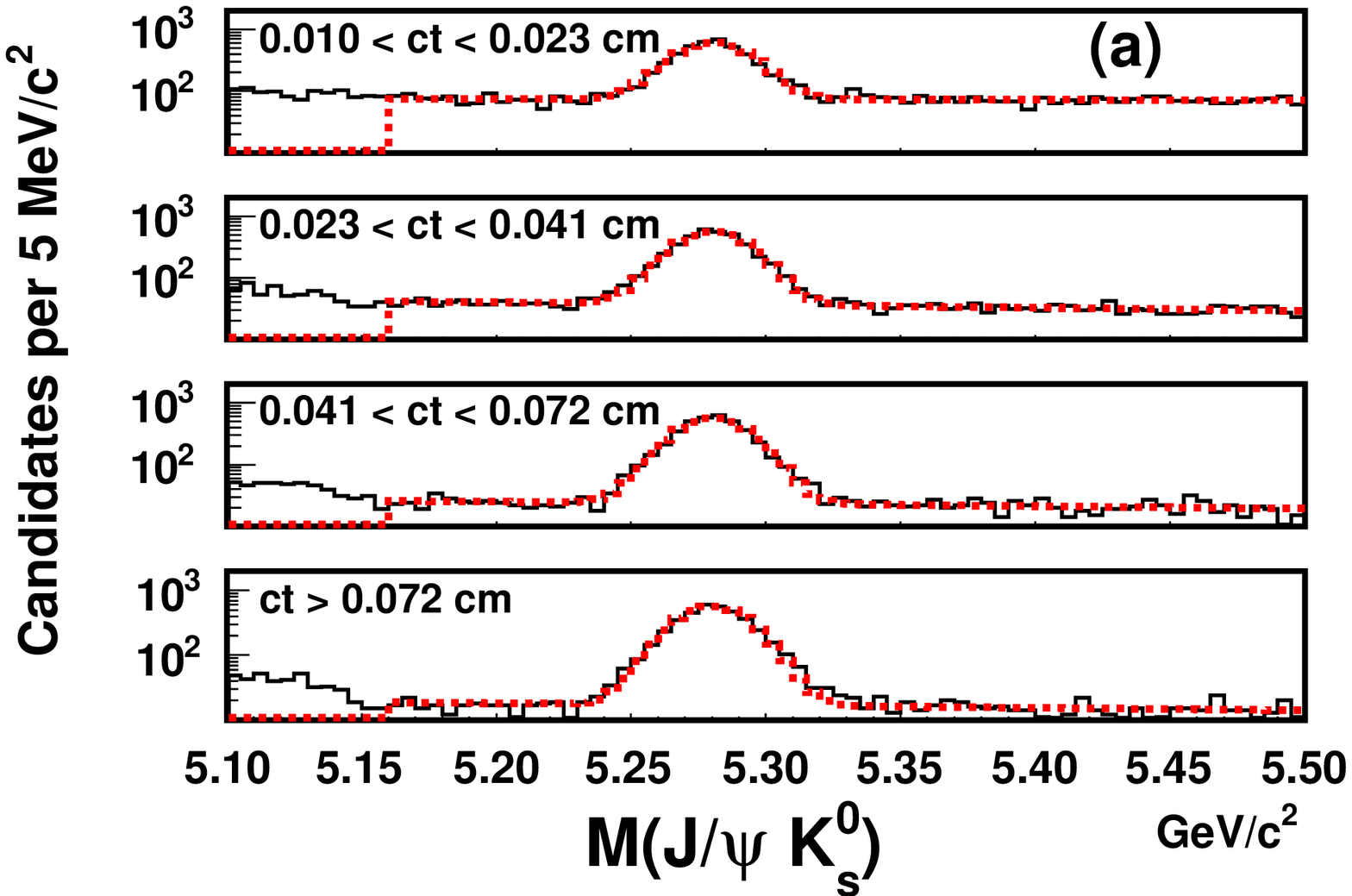,height=2.25in} 
\psfig{figure=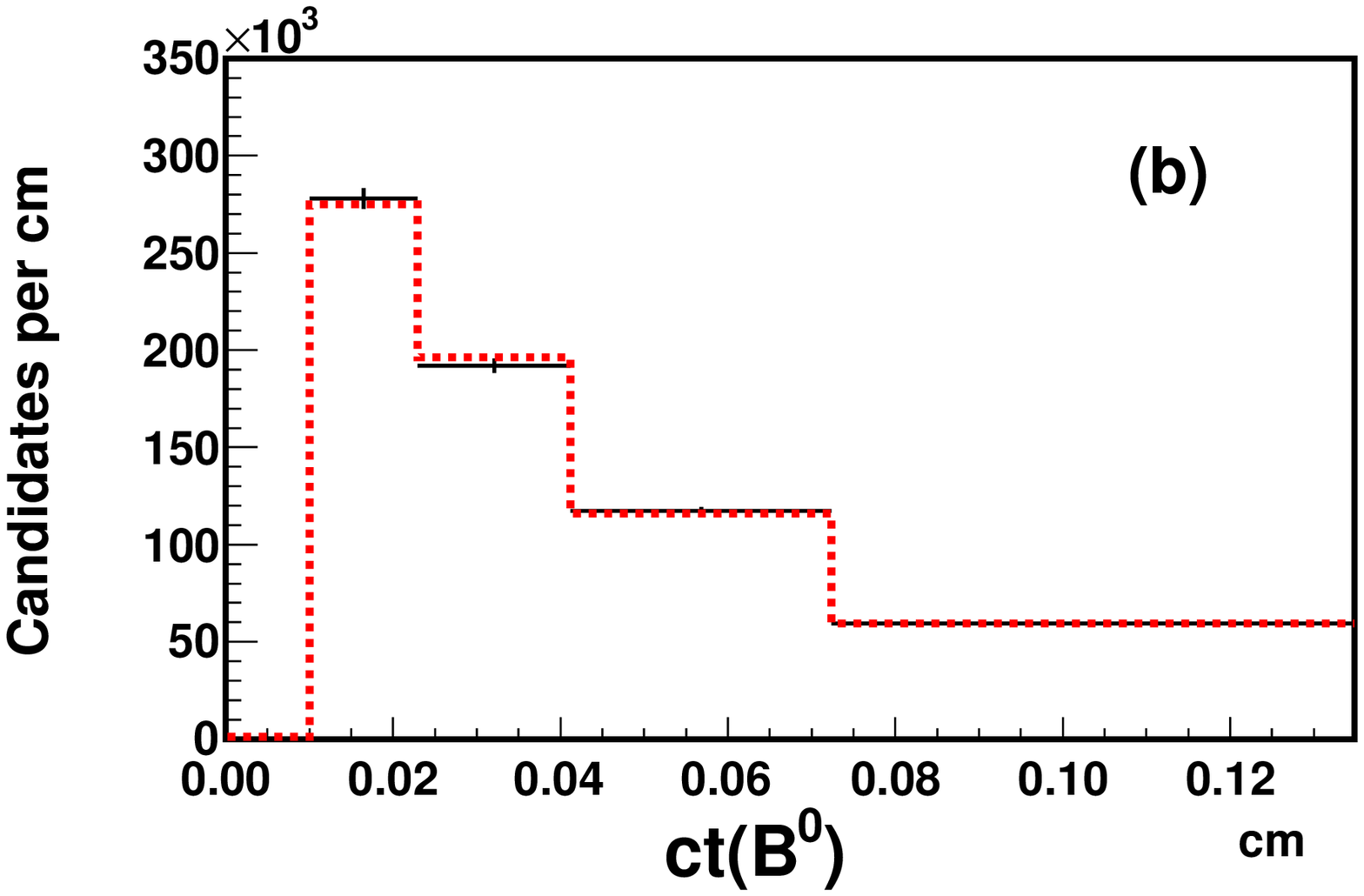,height=2.25in} 
\caption{Distributions of (a) $J/\psi \, K^{0}_S$ mass and
(b) $ct$ for $B^0$ reconstructed in the
 $B^0 \rightarrow J/\psi \, K^{0}_S$ decay.  
The mass and lifetime fits are 
overlaid in dashed red.
For display purposes, the upper limit of the $ct$ distribution is chosen 
to be 0.135 cm 
so that the displayed distribution contains 95\% of the candidates, based
on the initial lifetime estimate.
 \label{fig:fig_3}}
\end{figure}

The $\mu^+ \mu^-$ trigger data are well suited for use in particle property
measurements.  The trigger is insensitive to the decay time of any $b$ hadron,
so the corresponding samples are available for 
lifetime measurements without any trigger-induced
bias. 
This data sample provides the $B$ meson reference signals that are used
to verify mass and lifetime measurement techniques used in this analysis.
Mass measurements are obtained most directly from this fitting technique
by using all candidates with $ct > 100 \, \mu$m.  
The method
reduces to the unbinned mass distribution fit that is traditionally
used for mass measurements.  
Lifetime measurements are obtained by implementing the fit in
several decay-time bins as described previously.
As an example, 
the time-dependent mass distributions and decay-time distributions
for the $B^0 \rightarrow J/\psi K^0_S$ reference signal are shown in 
Fig.\ \ref{fig:fig_3}.
The results of the $B^+$ and $B^0$ measurements are listed in 
Table \ref{table:compare_ref} and discussed in Sec. \ref{sect:systematics}.

\begin{table*}[hbt]
\begin{center}
\caption{$B$ meson mass and $c\tau$ comparisons to known values \cite{PDG}.
Results from the entire data set (total) and the subset not included in the
world averages (new) are listed. Only statistical uncertainties are listed.
\protect \label{table:compare_ref}}
\vspace{3mm}
\begin{tabular}{cccccc}
\hline \hline
 Final state &  \multicolumn{2}{c}{Mass (MeV/$c^2$)}  & 
\multicolumn{2}{c}{$c\tau(\mu$m)} \\
\hline
\hline
 & Measured & Difference & Measured & Difference \\
  $J/\psi \, K^{+}   $ (total)       & $5278.75\pm0.06$ & $-0.5\pm0.2$  & 
$489.0\pm2.1$ & $-3.0\pm3.0$ \\
  $J/\psi \, K^{+}   $ (new) & $5278.74\pm0.08$ & $-0.5\pm0.2$ &  
$491.9\pm3.0$  & $-0.1\pm3.8$ \\
\hline
  $J/\psi \, K^{0*}  $ (total)       & $5279.01\pm0.11$ & $-0.5\pm0.2$  & 
$458.6\pm3.3$ & $\ \ 3.2\pm3.9$  \\
  $J/\psi \, K^{0*}  $ (new) & $5278.95\pm0.17$ & $-0.6\pm0.2$ &  
$458.4\pm4.7$ & $\ \ 3.0\pm5.1$  \\
\hline
 $J/\psi \, K^{0}_{s}$ (total)        & $5280.03\pm0.12$ & $\ \ 0.4\pm0.2$  & 
$458.6\pm4.2$ & $\ \ 3.2\pm4.6$ \\
 $J/\psi \, K^{0}_{s}$  (new) & $5280.09\pm0.18$ & $\ \ 0.5\pm0.2$ &  
$461.0\pm5.9$  & $\ \ 5.6\pm6.1$ \\
\hline
\hline
\end{tabular}
\end{center}
\end{table*}

\subsection{Masses of the $\Xi_c^0$ and $\Xi_c^+$ baryons \label{sect:Xic_mass}}
\begin{figure}[hbt]
\psfig{figure=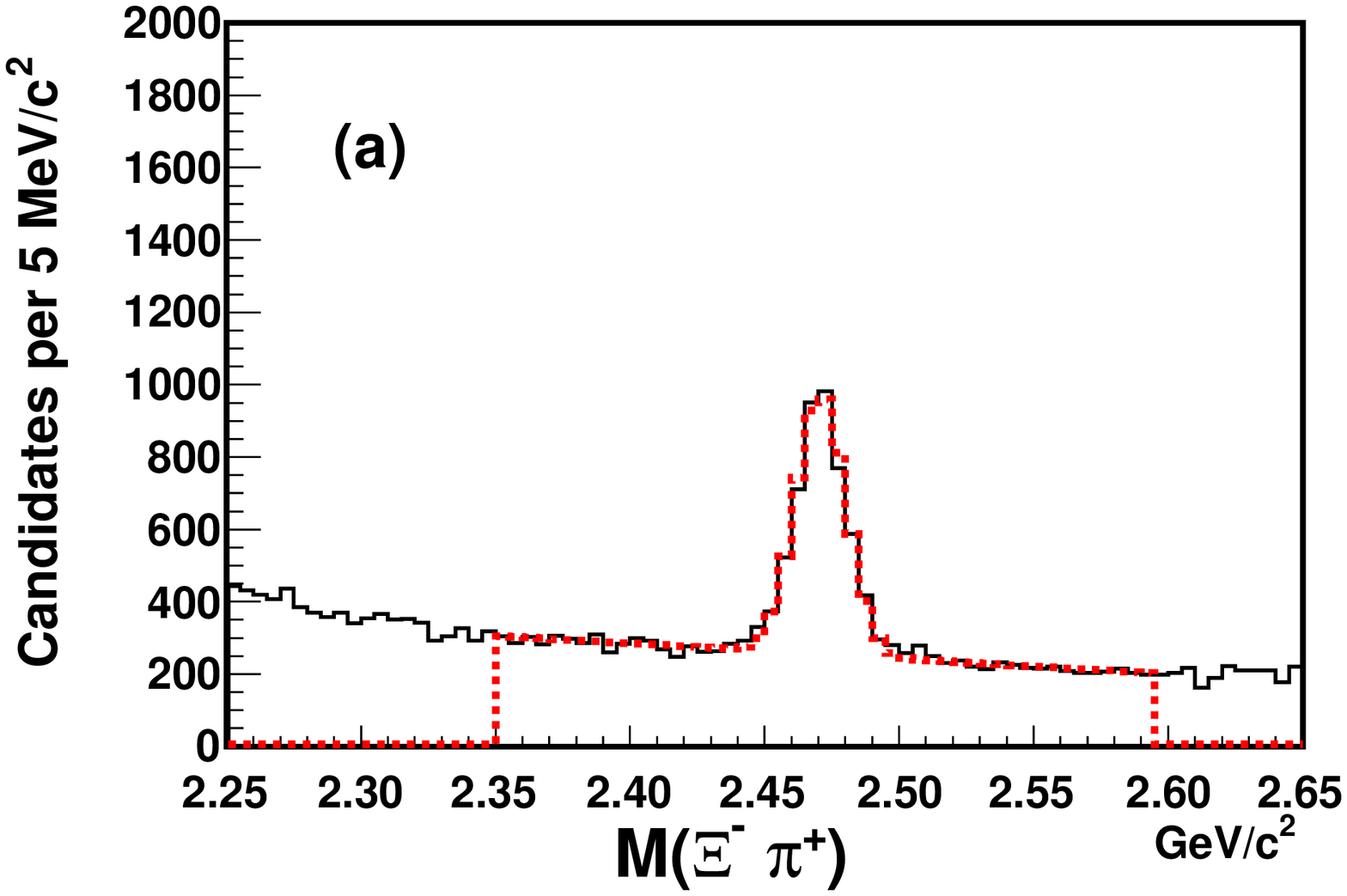,height=2.25in} 
\psfig{figure=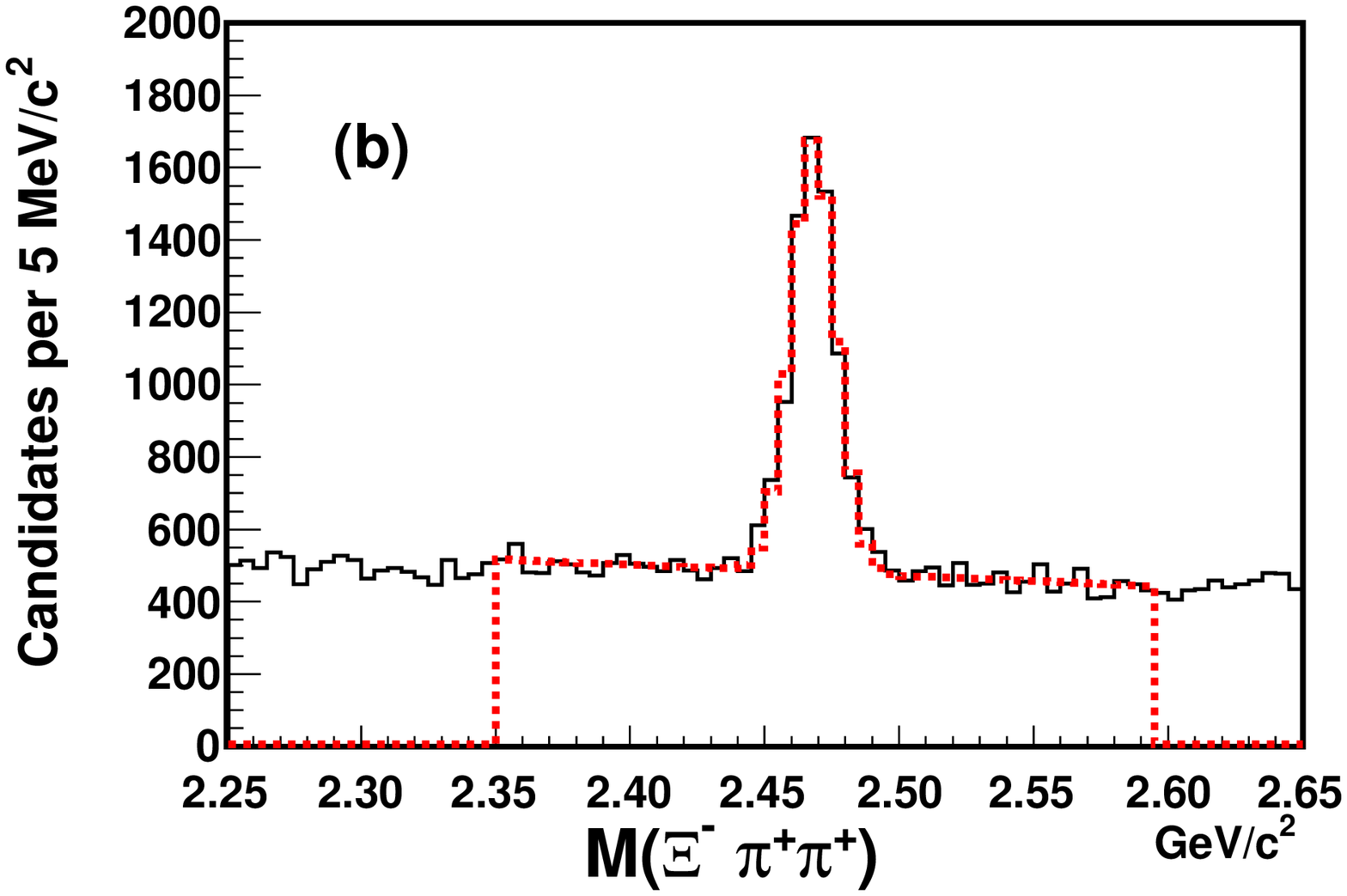,height=2.25in} 
\caption{Distribution of (a) $\Xi^- \, \pi^+$ and (b) $\Xi^- \, \pi^+ \, \pi^+$
mass used for the $\Xi_c$ mass measurements.
The fits are overlaid on the data in dashed red.
 \label{fig:fig_4}}
\end{figure}

The large samples
of $\Xi_c$ baryons 
in the full data set and the mass resolution available
from the tracking system allow precise $\Xi_c$ baryon mass measurements.  
The masses are obtained using  the unbinned
likelihood fit applied to all candidates with $ct > 100 \, \mu$m.
The $\Xi^- \pi^+$ and $\Xi^- \pi^+ \pi^+$
mass distributions along with projections of the fits are 
shown in Fig.\ \ref{fig:fig_4}.  
Results from the mass fits are listed in Table \ref{table:Lambdab_life}.

%
%
\begin{table}[hbt]
\begin{center}
\caption{Baryon mass and lifetime results.  
Only statistical uncertainties are listed .
\protect \label{table:Lambdab_life}}
\vspace{3mm}
\begin{tabular}{lccccc}
\hline \hline
 Final state & Mass (MeV/$c^2$) & $c\tau(\mu$m) & Yield \\
\hline
$\Xi_c^0 (\Xi^- \, \pi^+)$        & $2470.85\pm0.24$  & - & $3582\pm82$ \\ 
$\Xi_c^+ (\Xi^- \, \pi^+ \, \pi^+)$ & $2468.00\pm0.18$  & - & $5714\pm108$ \\ 
$\Lambda_b(J/\psi \, \Lambda)$    & $5620.15\pm0.31$ & $468.4\pm10.5$ & $2920\pm120$ \\ 
$\Xi_b^-(J/\psi \, \Xi^-)$  & $5793.2\pm1.9$ & $396\pm43$ & $112\pm19$ \\ 
$\Xi_b^-(\Xi^0_c \, \pi^-)$       & $5794.8\pm5.0$    & - & $33\pm6$ \\ 
$\Xi_b^0(\Xi^+_c \, \pi^-)$       & $5788.7\pm4.3$    & - & $62\pm9$ \\ 
 $\Omega_b^-(J/\psi \, \Omega^-)$ & $6050.0\pm4.1$ & $497^{+159}_{-119}$ & $22\pm6$ \\ 
$\Omega_b^-(\Omega_c^0 \, \pi^-)$ & $\ \ 6029\pm11$       & - & $\ \ 5.5\pm^{+2.5}_{-2.4}$ \\ 
\hline
\hline
\end{tabular}
\end{center}
\end{table}

\subsection{$\Lambda_b$ Measurements}

\begin{figure}[hbt]
\psfig{figure=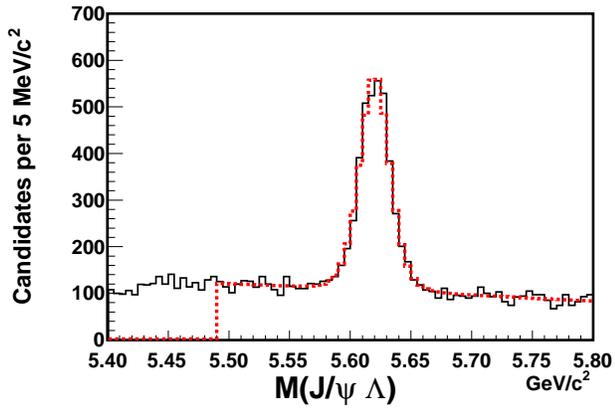,height=2.25in} 
\caption{Distribution of $J/\psi \, \Lambda$
mass used for the $\Lambda_b$ mass measurement.
The fit is overlaid on the data in dashed red.
 \label{fig:fig_5}}
\end{figure}

\begin{figure}[hbt]
\psfig{figure=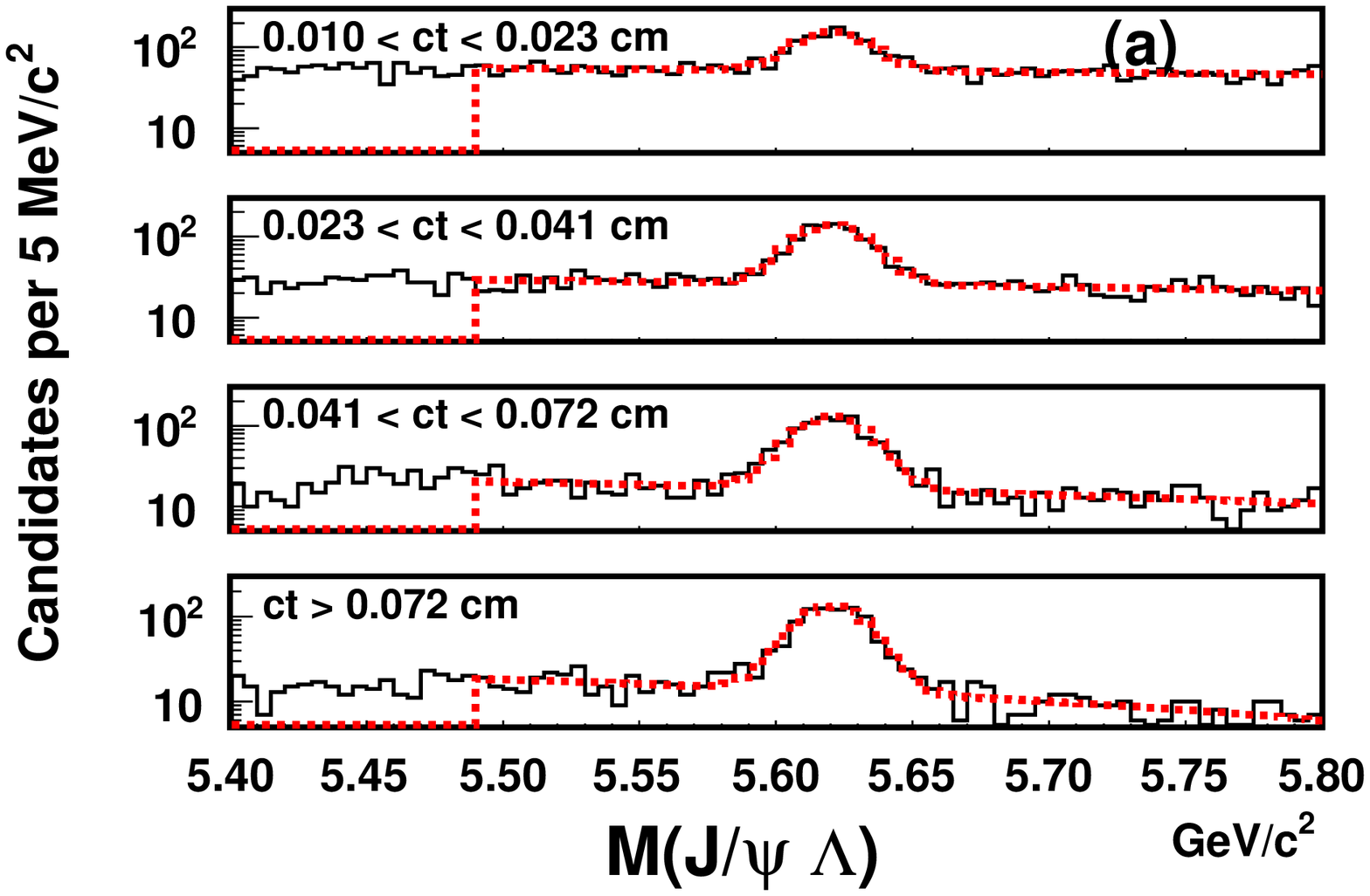,height=2.25in} 
\psfig{figure=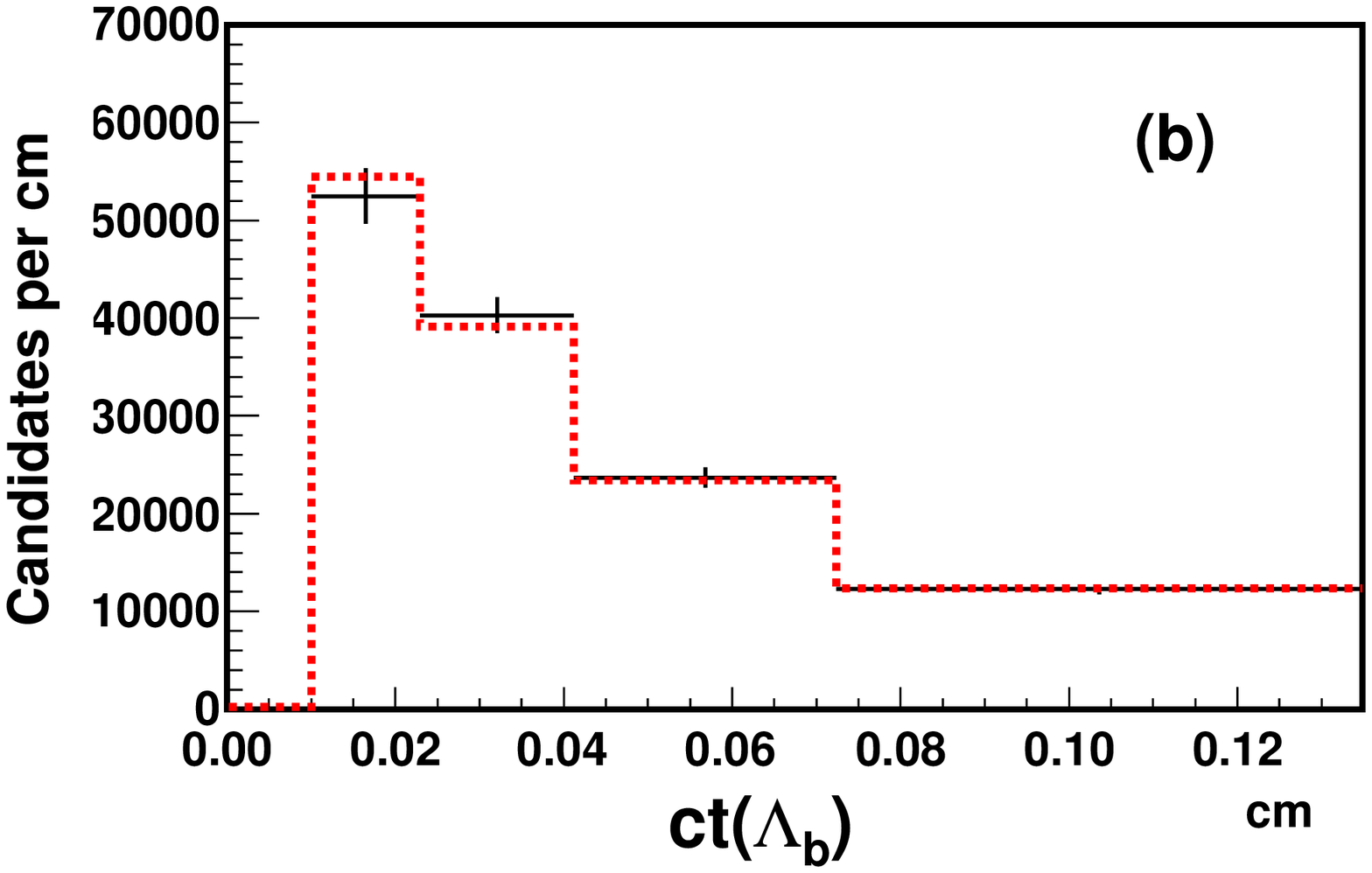,height=2.25in} 
\caption{Distribution of (a) $J/\psi \, \Lambda$ mass divided 
into four independent decay-time ranges and
(b) $ct$ of $\Lambda_b$ candidates 
used to calculate the lifetime.
The fits are overlaid on the data in dashed red.
 \label{fig:fig_6}}
\end{figure}

The approach to fitting the mass and lifetime of the $\Lambda_b$ is 
identical to that used for the meson reference signals.  
The mass distribution 
integrated in decay time and the 
projected fit are shown in Fig.\ \ref{fig:fig_5}.
The decay-time-dependent mass distributions and decay-time distribution
for the $\Lambda_b$ candidates are shown in 
Fig.\ \ref{fig:fig_6}.
Mass and lifetime results of the fits
are listed in Table \ref{table:Lambdab_life}.

The effect of reflections from the $B^0 \rightarrow J/\psi \, K^0_S$, 
$K^0_S \rightarrow \pi^+ \, \pi^-$ decays
is studied by recalculating the momenta of the
$J/\psi \Lambda$ candidates 
reconstructed under the $B^0 \rightarrow J/\psi \, K^0_S$ hypothesis.  
We find 
$420\pm29$ candidates consistent with
$B^0 \rightarrow J/\psi \, K^0_S$ decays within the $J/\psi \Lambda$
mass range used to fit the $\Lambda_b$.  This
$B^0$ background populates a portion of the $J/\psi \Lambda$ 
mass distribution 
that is systematically
lower than the $\Lambda_b$ mass.  The shape of the $B^0$ background
is parametrized and used as template for an additional
background component in an alternative fit to the $J/\psi \Lambda$
mass distribution.  The resulting $\Lambda_b$ yield shifts by 
approximately 1\% compared to the result obtained with the
simple linear background,
and is uniform over the time ranges.  The total shift is approximately
20\% of the statistical uncertainty in each time range, and fully 
correlated.  We conclude that any effect due to the
$B^0$ background is negligible with respect to other uncertainties.
Any systematic shift in the lifetime measurement due to $B^0$ background
treatment must be substantially less than 20\% of the statistical uncertainty.

\subsection{$\Xi_b^-$ measurements}
\begin{figure}[hbt]
\psfig{figure=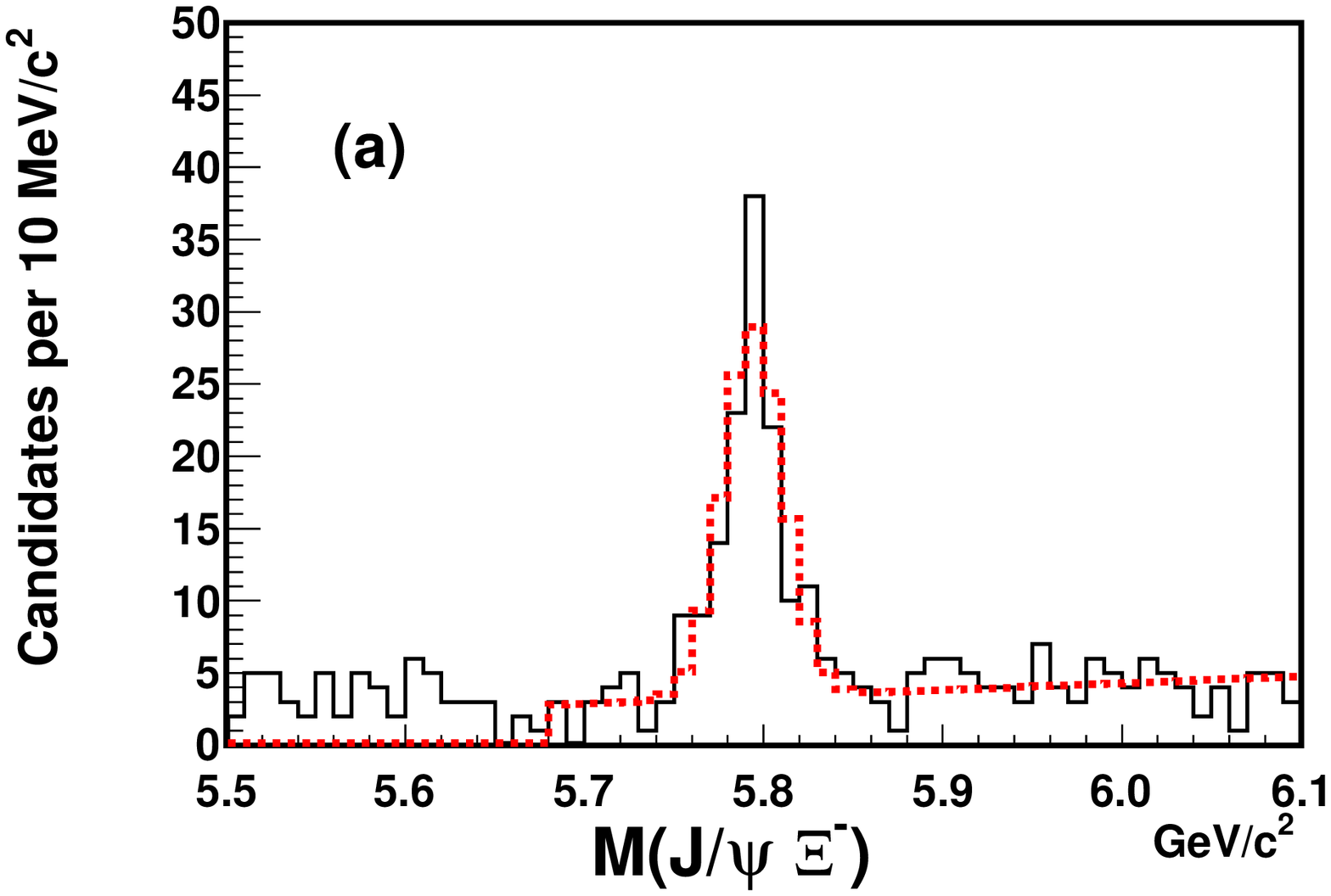,height=2.25in} 
\psfig{figure=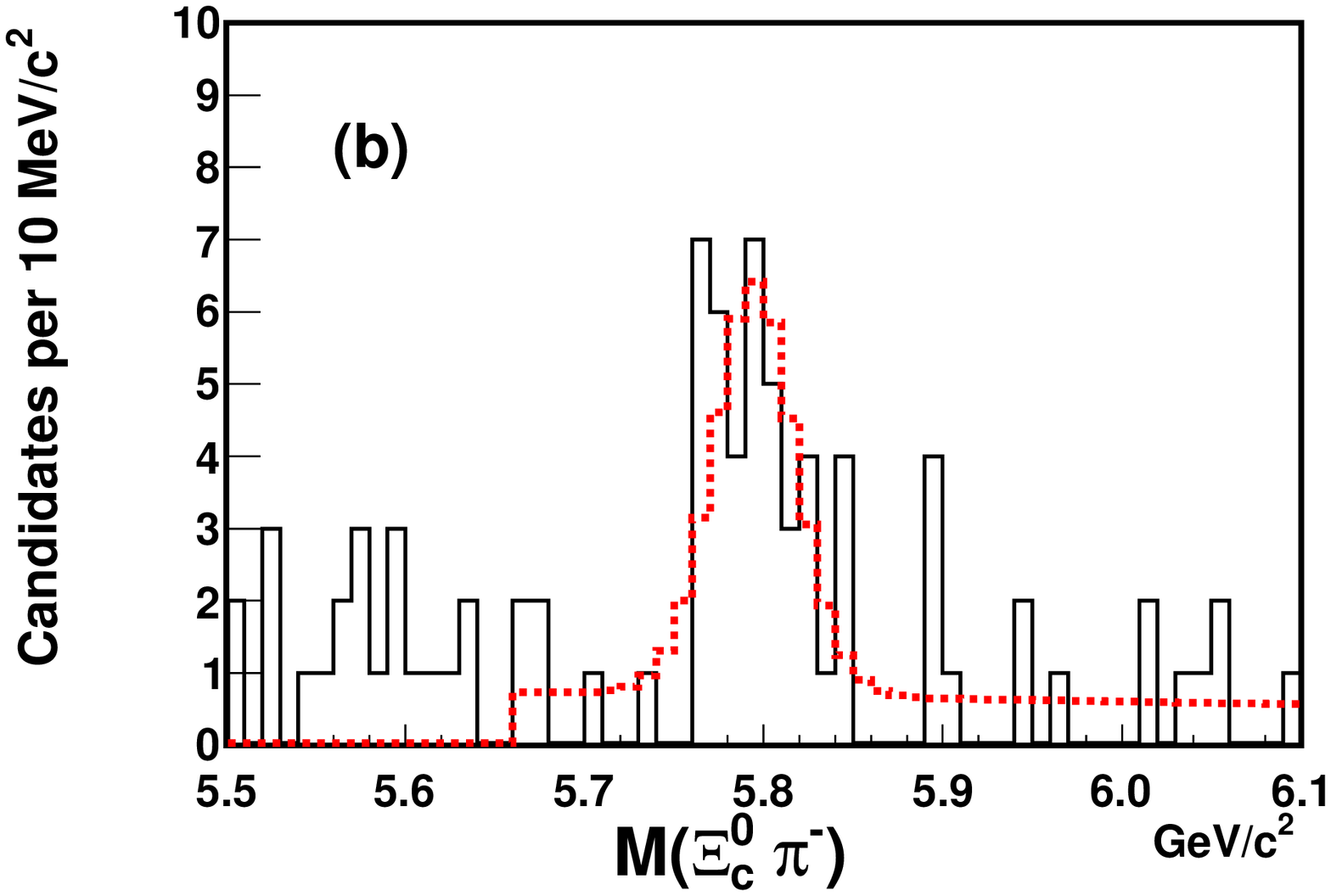,height=2.25in} 
\caption{Distribution of (a) the $J/\psi \, \Xi^-$ and
(b)  $\Xi_c^0 \, \pi^-$ 
mass used for the $\Xi^-_b$ mass measurements.
The fits are overlaid on the data in dashed red.
 \label{fig:fig_7}}
\end{figure}

\begin{figure}[hbt]
\psfig{figure=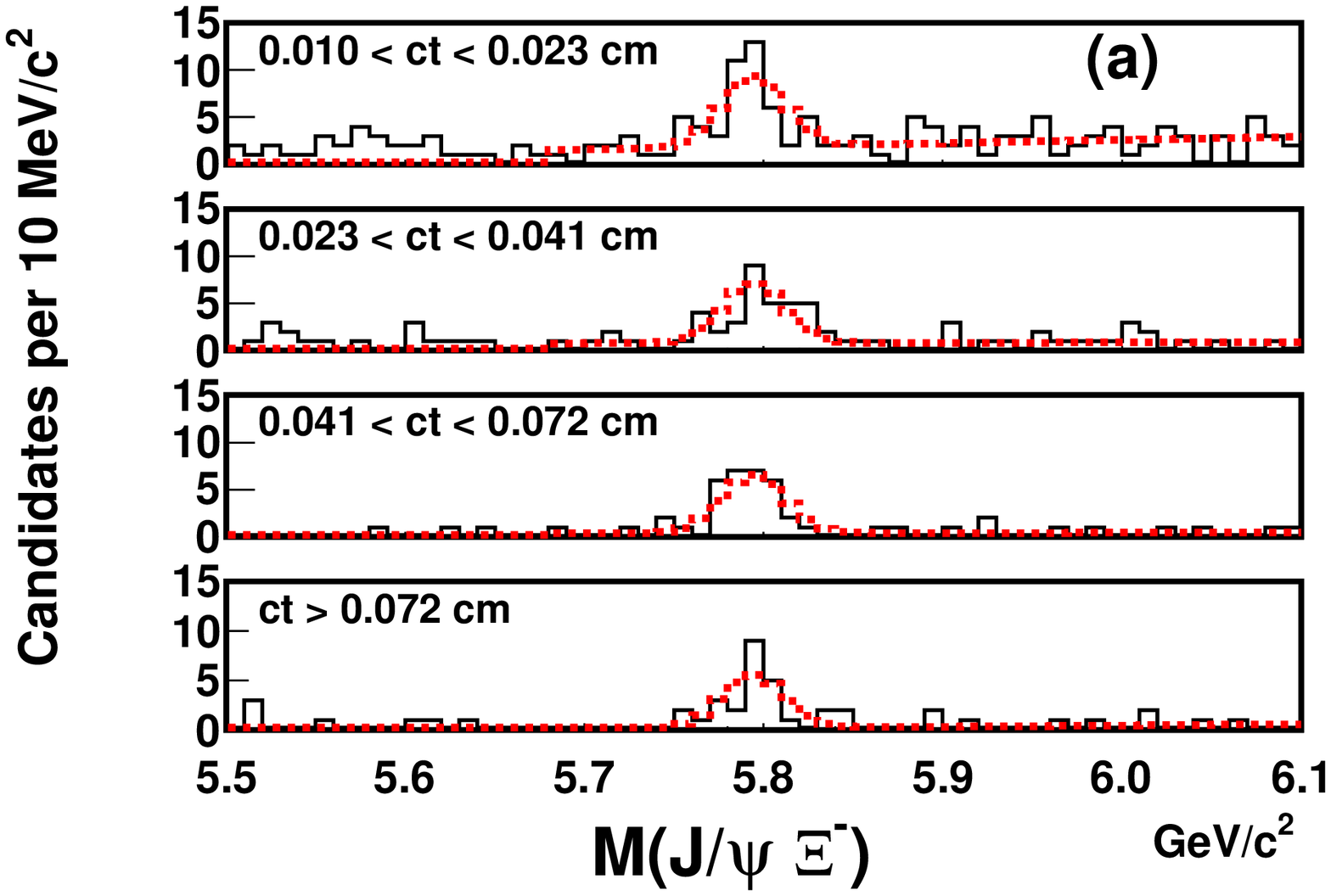,height=2.25in} 
\psfig{figure=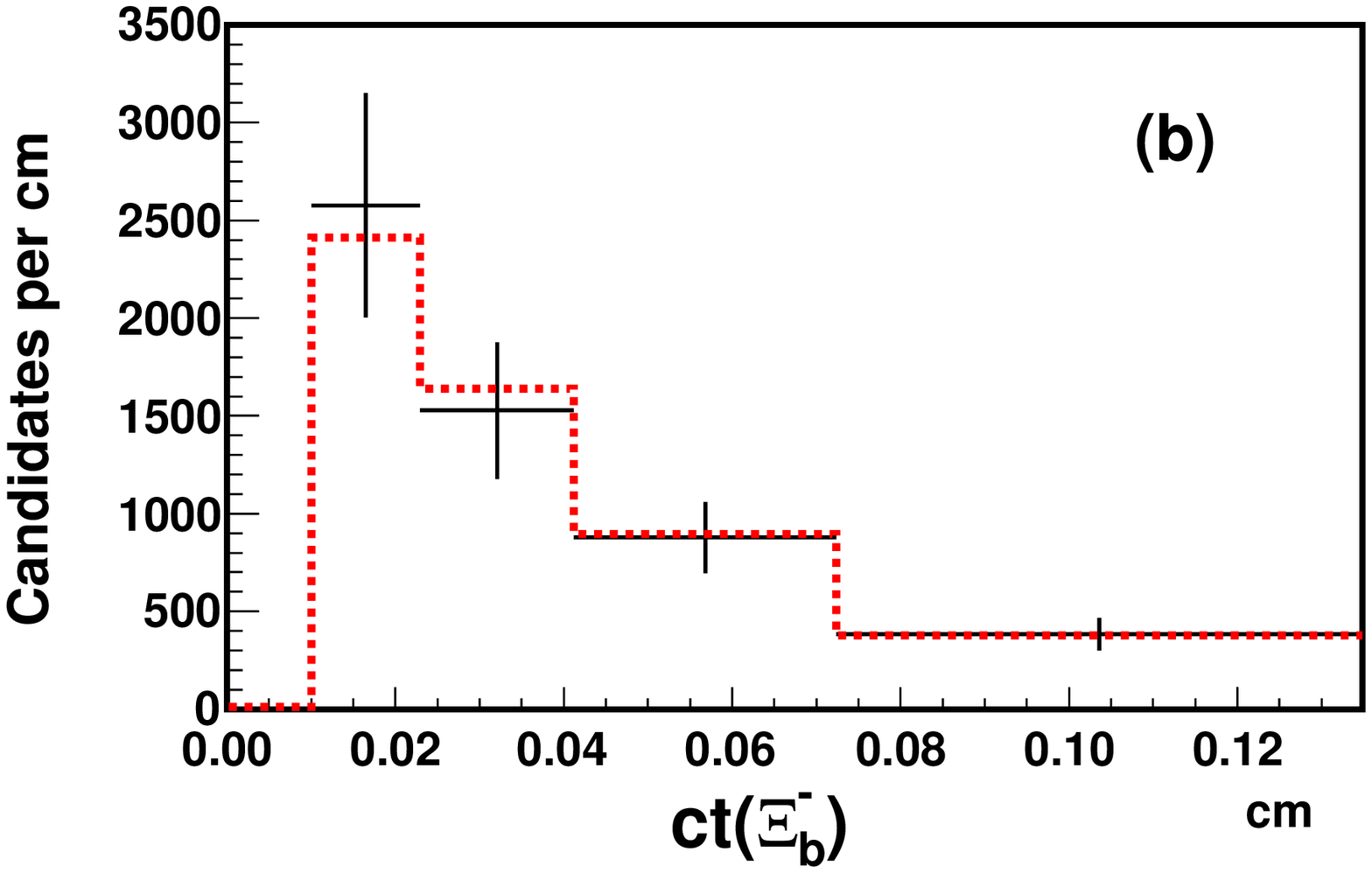,height=2.25in} 
\caption{Distribution of (a) $J/\psi \, \Xi^-$ mass divided 
into four independent decay-time ranges and
(b) $ct$ of $\Xi_b^-$ candidates 
used to calculate the lifetime.
The fits are overlaid on the data in dashed red.
 \label{fig:fig_8}}
\end{figure}

The approach to fitting the mass and lifetime of the $\Xi_b^-$ is 
identical to that used for the meson reference signals.  
The mass distributions for the $J/\psi \, \Xi^-$  and 
$\Xi_c^0 \, \pi^-$ combinations integrated in decay time and the 
projected fits are shown in Fig.\ \ref{fig:fig_7}.
The time-dependent mass distributions and decay-time distribution
for the $\Xi_b^- \rightarrow J/\psi \, \Xi^-$ channel are shown in 
Fig.\ \ref{fig:fig_8}.
Mass and lifetime results of the fits
are listed in Table \ref{table:Lambdab_life}.

\subsection{$\Xi_b^0$ measurements}
\begin{figure}[hbt]
\psfig{figure=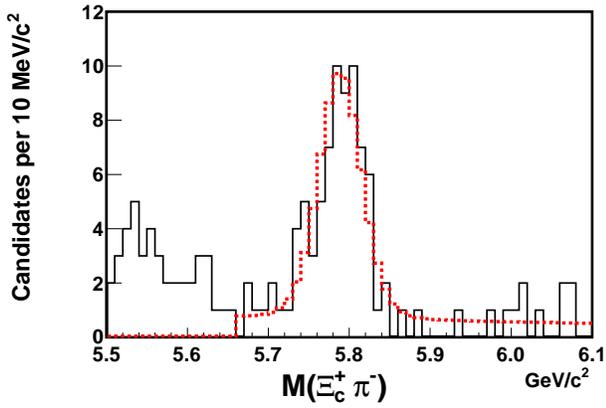,height=2.25in} 
\caption{Distribution of  $\Xi_c^+ \, \pi^-$
mass used for the $\Xi^0_b$ mass measurement.
The fit is overlaid on the data in dashed red.
 \label{fig:fig_9}}
\end{figure}

The process $\Xi_b^0 \rightarrow J/\psi \, \Xi^0$ 
is expected to occur, in analogy to $\Xi_b^- \rightarrow J/\psi \, \Xi^-$.
However, this process requires the accurate reconstruction of a 
low-momentum $\pi^0$, so it is outside the sensitivity of this experiment.
Consequently, we are limited to a $\Xi_b^0$ mass measurement in the
$\Xi_b^0 \rightarrow \Xi_c^+ \, \pi^-$ channel.
The $\Xi_c^+ \, \pi^-$ mass distribution and the projection of the fit
overlaid on the data are shown in Fig.\ \ref{fig:fig_9}, and the fit result
is listed in Table \ref{table:Lambdab_life}.

\subsection{$\Omega_b^-$ measurements}
\begin{figure}[hbt]
\psfig{figure=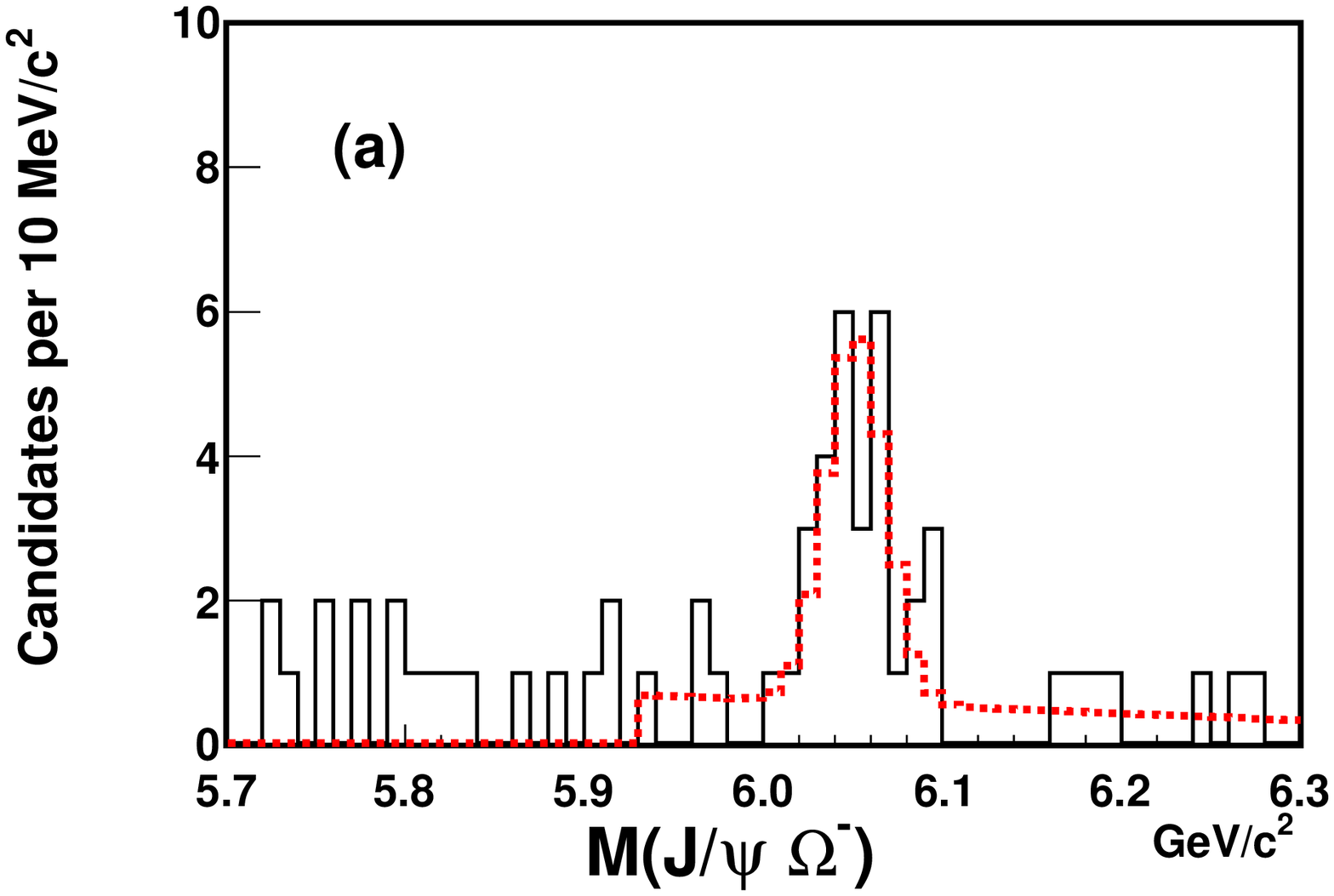,height=2.25in} 
\psfig{figure=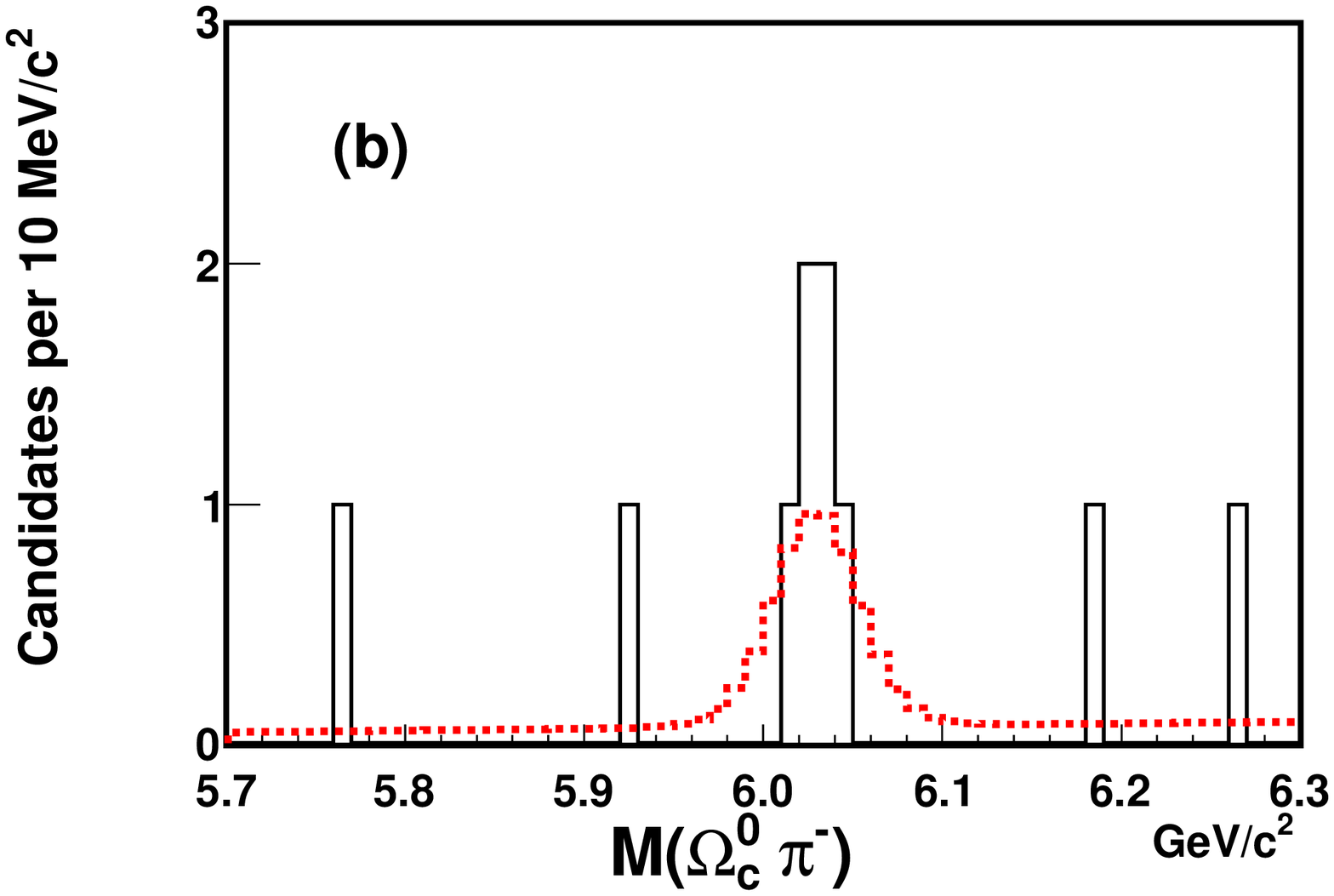,height=2.25in} 
\caption{Distribution of (a) $J/\psi \, \Omega^-$ and
(b) $\Omega_c^0 \, \pi^-$
mass used for the $\Omega^-_b$ mass measurement.
The fits are overlaid on the data in dashed red.
 \label{fig:fig_10}}
\end{figure}

\begin{figure}[hbt]
\psfig{figure=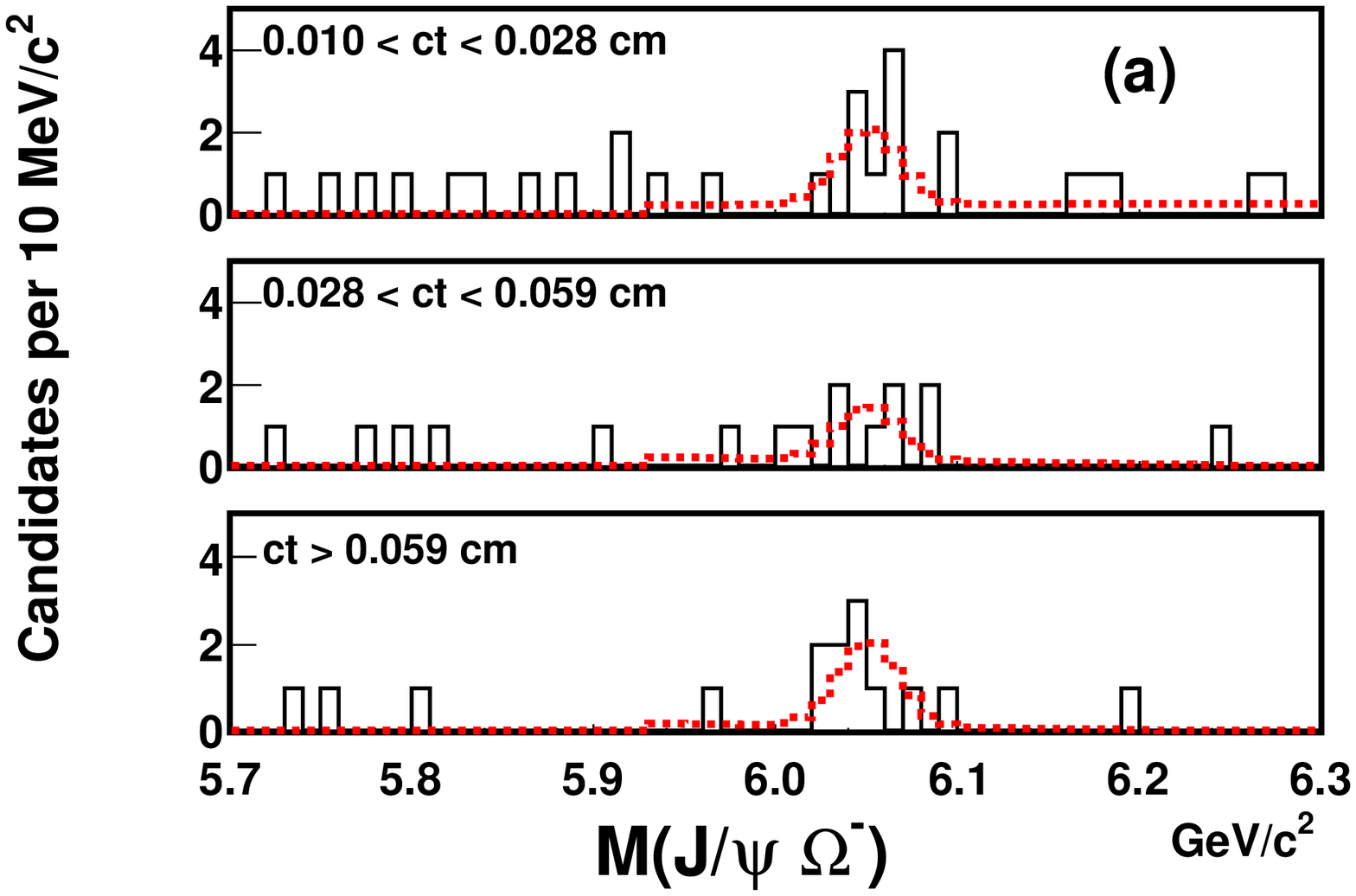,height=2.25in} 
\psfig{figure=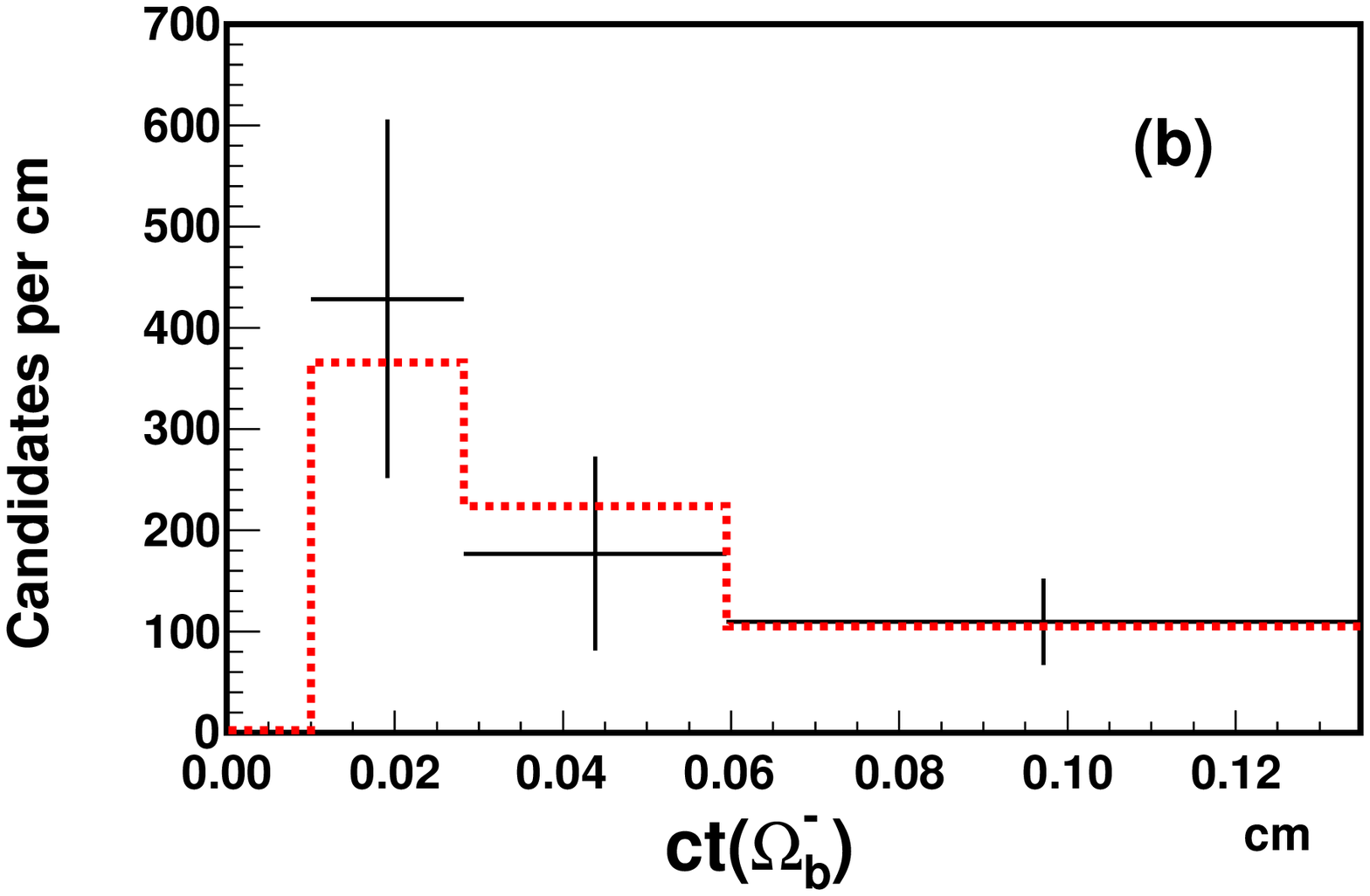,height=2.25in} 
\caption{Distribution of (a) $J/\psi \, \Omega^-$ mass 
divided into three independent decay-time ranges and
(b) $ct$ of $\Omega_b^-$ candidates 
used to calculate the lifetime.
The fits are overlaid on the data in dashed red.
 \label{fig:fig_11}}
\end{figure}

The approach of fitting the mass and lifetime of the $\Omega_b^-$ is
identical to that used for the meson reference signals, 
with the exception that   
only three decay-time ranges are used in the
lifetime calculation due to the small sample of candidates.
The mass distributions for the $J/\psi \, \Omega^-$ and 
$\Omega_c^0 \, \pi^-$ combinations integrated in decay time and the 
projected fits are shown in Fig.\ \ref{fig:fig_10}.
The time-dependent mass distributions and decay-time distribution
for the $\Omega_b^- \rightarrow J/\psi \, \Omega^-$ channel are shown in 
Fig.\ \ref{fig:fig_11}.  
The mass resolution terms $s_0$ are fixed to the values obtained 
in the analogous channels used in the $\Xi^-_b$ fits.
The results of the mass and lifetime fits are listed in 
Table \ref{table:Lambdab_life}.

\subsection{Systematic uncertainties
 \label{sect:systematics}}
The systematic uncertainties on the mass measurements
reported here are similar to those
obtained for other $b$ hadrons in previous CDF II analyses.
The mass scale uncertainty is taken from
earlier work \cite{CDF_B_mass}.  Here, the $J/\psi$, $\psi(2S)$ and 
$\Upsilon$ decays,
reconstructed in dimuon final states,
were used to set the mass scale.  The 
differences of the measured masses from the 
true masses are parametrized as functions of the 
total kinetic energy in these decays and are then used to obtain the
mass scale uncertainties listed in Table \ref{table:systematics}.
The effect of the mass-resolution model on the mass-uncertainty scale
is tested in several variations of the fits on the $B$ meson 
and $\Xi_c$ baryon signals.  These variations
 indicate that the choice of resolution model
can affect the resulting mass measurement by 0.05 -- 0.1 MeV/$c^2$.  
The effect of the tracking-system material on the
mass scale is tested by examining 
the mass of the $B^0$ reconstructed in the
$J/\psi \, K^0_S$ channel as a function of the $K^0_S$
decay point.  A systematic shift of $1.1\pm0.5$ MeV/$c^2$ 
is found for combinations where the $K^0_S$ decay point
is outside the silicon system.  
In order to determine the sensitivity of $b$-baryon reconstruction
to the tracking-system material,
we determine the fraction of $b$-baryon candidates 
whose $\Lambda, \Xi^-$, or
$\Omega^-$ decay outside the silicon system by using the data for 
the $\Xi_c$ and 
$\Lambda_b$ baryons and simulation for the $\Xi_b$ and $\Omega_b^-$ baryons.
This fraction of the shift observed in $B^0$ decays is taken for a systematic
uncertainty on the mass due to material description of the detector.
Masses of the $p, \, \pi^-, \, K^-,$ and $\Lambda$ are sufficiently 
well determined that their contribution to the systematic uncertainty
is insignificant.  
These effects are listed in Table \ref{table:systematics}, where they are
combined in quadrature to obtain the total systematic uncertainties.  
\begin{table*}[hbt]
\begin{center}
\caption{Contributions to the systematic uncertainty on the 
mass measurements.
\protect\label{table:systematics}}
\vspace{3mm}
\begin{tabular}{lcccccccccc}
\hline \hline
 Source & \multicolumn{9}{c}{Uncertainty (MeV/$c^2$)} \\
\hline
 & $\Xi_c^0$  & $\Xi_c^+$ &  \multicolumn{2}{c}{$B^{0}$} & $\Lambda_b$ &  
\multicolumn{2}{c}{$\Xi_b^{-}$} & 
$\Xi_b^{0}$ & \multicolumn{2}{c}{$\Omega_b^{-}$} \\  
\hline
 & & & $J/\psi K^{0*}$ & $J/\psi K^{0}_S$ & & 
$J/\psi \Xi^-$ & $\Xi_c^0 \, \pi^-$ & & 
$J/\psi \Omega^-$ & $\Omega_c^0 \, \pi^-$ \\  
Mom. scale    & 0.35 & 0.35 & \ 0.42 & \ 0.45 & \ 0.40 & \ 0.40 & \ 0.50 & \ \ 0.40 & \ 0.40 &  \ 0.55 \\
Reso. model & 0.05 & 0.05 & 0.1 & 0.1 & 0.1 & 0.1 & 0.1 & 0.1 & 0.1 & 0.1 \\
Material & 0.38 & 0.38 & 0.0 & \ 0.25 & \ 0.21 & \ 0.47 & \ \, 1.16 & \ \, 1.15 & \ 0.38 & \ 0.94 \\
\hline
Total    & 0.55 & 0.51 & \ 0.43 & \ 0.53 & \ 0.47 & 0.6 & \ 1.4 & \ 1.4 & 0.6 & 1.2 \\
\hline
\hline
\end{tabular}
\end{center}
\end{table*}

The $B$ mesons reconstructed in the $J/\psi$ sample serve as 
a precision reference sample to support the evaluation of
the systematic uncertainties.
The mass and lifetime results obtained for the $B^+$ and $B^0$ are listed in 
Table \ref{table:compare_ref}.
Comparisons between the measurements and the known values \cite{PDG} 
are listed.
The known values contain contributions from a subset of the CDF Run II 
data \cite{CDF_B_mass,CDF_B_life}.  
Consequently, values are given for the full data set and the data
taken since the previous measurement \cite{CDF_Omega_b}.  
We find the more recent data to be completely consistent with our earlier
measurements, indicating that no significant degradation of the tracking 
resolution occurred.
A comparison of the results in reference signals 
with the known values demonstrates that the mass measurements are
well understood.

The masses of the $\Xi_b^-$ and $\Omega_b^-$ baryons
are each measured in two different
final states.  The mass results are combined to provide a single 
measurement following Ref. \cite{Lyons}.  These
combined results, and the mass results for the other baryons, 
are listed in Table \ref{table:final_prop}.
The momentum scale uncertainty cancels
in measurements of the mass differences between the 
$\Xi_c^0$ and $\Xi_c^+$, and the $\Xi_b^-$ and $\Xi_b^0$ baryons.  
The estimates for these isospin splittings are also listed in 
Table~\ref{table:final_prop}.

\begin{table}[hbt]
\begin{center}
\caption{$\Xi_c$ and $b$-baryon mass results.  The first uncertainty
listed is statistical and the second is systematic.
\protect \label{table:final_prop}}
\vspace{3mm}
\begin{tabular}{cccc}
\hline \hline
Baryon &  Mass (MeV/$c^2$) \\
\hline
$\Xi_c^{0}$  & $2470.85\pm0.24\pm0.55$  \\
$\Xi_c^{+}$  & $2468.00\pm0.18\pm0.51$  \\
$\Lambda_b$  & $5620.15\pm0.31\pm0.47$  \\
$\Xi_b^-$    & $5793.4\pm1.8 \ \, \pm0.7$  \\
$\Xi_b^0$    & $5788.7\pm4.3 \ \, \pm1.4$  \\
$\Omega_b^-$ & $6047.5\pm3.8 \ \, \pm0.6$  \\
\hline
$M(\Xi_c^0)-M(\Xi_c^+)$ & $\ \ \ \ 2.85\pm0.30\pm0.04$ \\
$M(\Xi_b^-)-M(\Xi_b^0)$ & $\ \ \ 4.7 \, \pm4.7 \ \, \pm0.7$\\
\hline
\hline
\end{tabular}
\end{center}
\end{table}

\begin{figure}[hbt]
\psfig{figure=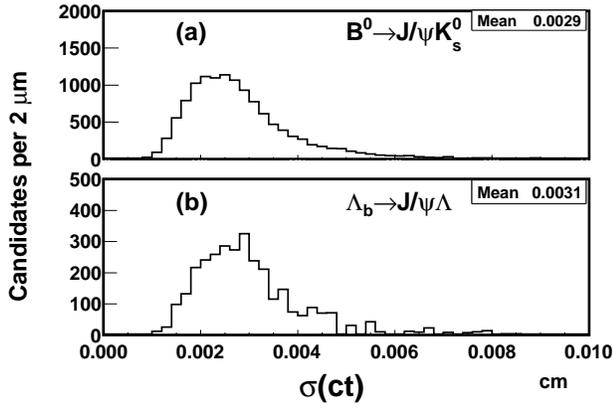,height=2.25in} 
\caption{Distribution of $\sigma_{ct}$ for
(a) $B^0 \rightarrow J/\psi \, K^0_S$ candidates and
(b) $\Lambda_b \rightarrow J/\psi \, \Lambda$ candidates.
 \label{fig:fig_12}}
\end{figure}

The lifetime fit is repeated on the reference signals to determine
the sensitivity of the technique to the input parameters chosen for the fit.  
The decay-time uncertainty $\sigma_{ct}$ is 
shown in Fig.\ \ref{fig:fig_12} for the 
$B^0 \rightarrow J/\psi \, K^0_S$  sample, where the background contribution
is removed by subtracting mass sideband uncertainties.  
If this uncertainty is varied between $15$ and $45 \, \mu$m,
the results of the lifetime fit are found to have
a relative variation of less than $10^{-3}$.
Variations in the number of decay-time bins have similar impact.  

Systematic uncertainties on the lifetime measurements of the $b$ baryons
 are identical to those of the $B$ mesons.  The $B$ meson 
reference signals all have lifetime results that are within 1\% of 
their known values, as is shown in Table \ref{table:compare_ref}.
If we use only the recent data for the $B^0 \rightarrow J/\psi \, K^0_S$
process, we find complete consistency with the known lifetime
within $\pm6 \, \mu$m, or 1.3\%.  This is taken as 
the systematic uncertainty for all $b$-baryon lifetime measurements.

\begin{table}[hbt]
\begin{center}
\caption{$b$-baryon lifetime results.   The first uncertainty
listed is statistical and the second is systematic.
\protect \label{table:final_life}}
\vspace{3mm}
\begin{tabular}{cccc}
\hline \hline
Baryon &  lifetime (ps) \\
\hline
$\Lambda_b$  & $1.565 \, \pm0.035\pm0.020$ \\
$\Xi_b^- $   & $1.36 \, \pm0.15 \ \ \pm0.02$ \\
$\Omega_b^-$ & $1.66^{+0.53}_{-0.40} \ \  \ \  \ \ \pm0.02$ \\
\hline
\hline
\end{tabular}
\end{center}
\end{table}

\section{Final Results \label{sect:results}}
Final results for the properties of the $b$ baryons are listed in 
Tables \ref{table:final_prop} and \ref{table:final_life}.
The measurements of the masses of the $\Xi_c$ baryons are competitive with the 
world averages, and consistent with them \cite{PDG}.  
The isospin splitting 
of the states is also comparable to the world average.  
These measurements serve to improve our overall knowledge 
of heavy baryon dynamics.   
Theoretical calculations of the $\Xi_c$ baryon masses
are not
as precise as the current measurements, so these results serve to 
constrain the models considered for heavy baryon mass 
predictions \cite{Jenkins,Karliner}.
As with the charmed baryons, the $\Lambda_b$
mass is now known with high precision.
All other $b$ baryons are currently only seen in small samples,
so the measurements are limited by the sample size.  
The mass obtained for the $\Xi_b^0$ confirms our earlier 
observation \cite{CDF_Xib0}
and provides a small improvement to our unique measurement of the 
isospin splitting in the $\Xi_b$ system, which 
is consistent with the prediction of Ref.~\cite{Karliner}.  
The present measurement of the mass of the $\Omega_b^-$
provides further support to our first result \cite{CDF_Omega_b} and 
is inconsistent with the measurement associated with the
first observation of this particle \cite{D0_Omega_b}.

The precision of the measurement of the lifetime of the $\Lambda_b$ 
baryon is  comparable to that of the
$B^0$ meson.  By combining the two reference
measurements of the $B^0$
and retaining the full systematic uncertainty found in 
Sect. \ref{sect:systematics}, we obtain
$\tau(\Lambda_b)/\tau(B^0)=
 1.021\pm0.024(\textrm{stat})\pm0.013(\textrm{syst})$,
which is more consistent with the predicted values \cite{Neubert,Rosner} 
than earlier measurements.
The lifetime measurements of the $\Xi_b^-$
and $\Omega_b^-$ baryons are unique and limited by the size of the samples.
The values obtained appear to be typical of other $b$ hadrons.

\section{Conclusions \label{sect:Conclusions}}
In conclusion, the CDF Run II data set is analyzed to identify the 
largest possible low-background
sample of $\Xi_c$ and $b$-baryon ground states.  
The mass and lifetime properties
of these particles are measured, and the results compared to  
precisely measured quantities for $B$ mesons obtained in similar final states.
The mass and isospin splitting of the $\Xi_c$ system are measured with  
precisions that are comparable to the world averages.
The first evidence for the process 
$\Omega_b^- \rightarrow \Omega_c^0 \, \pi^-$ is shown.
The masses and lifetimes of the $\Lambda_b$, $\Xi_b^-$ and $\Omega_b^-$
baryons are
measured and are found consistent with LHCb determinations
\cite{LHCb_Xib,LHCb_Lambdab_life,LHCb_Xib0}.
The  isospin splitting of the 
$\Xi_b$ system is unique to this experiment and is updated with the
final data set.
These results supersede previous measurements, 
which were obtained using a subset of these data \cite{CDF_Omega_b,CDF_Xib0}.

We thank the Fermilab staff and the technical staffs of the
participating institutions for their vital contributions. This work
was supported by the U.S. Department of Energy and National Science
Foundation; the Italian Istituto Nazionale di Fisica Nucleare; the
Ministry of Education, Culture, Sports, Science and Technology of
Japan; the Natural Sciences and Engineering Research Council of
Canada; the National Science Council of the Republic of China; the
Swiss National Science Foundation; the A.P. Sloan Foundation; the
Bundesministerium f\"ur Bildung und Forschung, Germany; the Korean
World Class University Program, the National Research Foundation of
Korea; the Science and Technology Facilities Council and the Royal
Society, United Kingdom; the Russian Foundation for Basic Research;
the Ministerio de Ciencia e Innovaci\'{o}n, and Programa
Consolider-Ingenio 2010, Spain; the Slovak R\&D Agency; the Academy
of Finland; the Australian Research Council (ARC); and the EU community
Marie Curie Fellowship Contract No. 302103.

\end{document}